\documentclass[english,11pt,a4paper]{article}
\usepackage[T1]{fontenc}
\usepackage[latin9]{inputenc}
\usepackage{geometry}[25mm]
\usepackage{graphicx}
\usepackage{tikz}
\usepackage{color}
\usepackage{babel}
\usepackage{amsmath}
\usepackage{amsthm}
\usepackage{amssymb}
\usepackage{dsfont}

\usepackage[unicode=true,pdfusetitle,
 bookmarks=true,bookmarksnumbered=false,bookmarksopen=false,
 breaklinks=false,pdfborder={0 0 0},pdfborderstyle={},backref=false,colorlinks=true]
 {hyperref}
\hypersetup{
 linkcolor=vblue, citecolor=vblue}

\makeatletter

\numberwithin{equation}{section}
\theoremstyle{plain}
\newtheorem{prop}{\protect\propositionname}
\theoremstyle{plain}
\newtheorem{thm}{\protect\theoremname}
\theoremstyle{plain}
\newtheorem{lem}{\protect\lemmaname}
\theoremstyle{plain}

\theoremstyle{plain}
\newtheorem*{prop*}{\protect\propositionname}

\DeclareMathOperator{\kin}{kin}
\DeclareMathOperator{\inter}{int}

\DeclareMathOperator{\FS}{FS}

\definecolor{vblue}{RGB}{0,0,255}

\makeatother

\providecommand{\corollaryname}{Corollary}
\providecommand{\lemmaname}{Lemma}
\providecommand{\propositionname}{Proposition}
\providecommand{\theoremname}{Theorem}

\begin{document}
\title{On the Effective Quasi-Bosonic Hamiltonian of the Electron Gas: Collective Excitations and Plasmon Modes\footnote{This
article belongs to the themed collection: Mathematical Physics and Numerical
Simulation of Many-Particle Systems; V. Bach and L. Delle Site (eds.)}}
\author{Martin Ravn Christiansen, Christian Hainzl, Phan Th\`anh Nam\\
\\
{\footnotesize{}Department of Mathematics, Ludwig Maximilian University
of Munich, Germany}\\
{\footnotesize{}Emails: christiansen@math.lmu.de, hainzl@math.lmu.de,
nam@math.lmu.de}}
\maketitle
\begin{abstract}

We consider an effective quasi-bosonic Hamiltonian of the electron gas which emerges naturally from the random phase approximation and describes the collective excitations of the gas. By a rigorous argument, we explain how the plasmon modes can be interpreted as a special class of approximate eigenstates of this model.

\end{abstract}
\tableofcontents{}

\section{Introduction}

In a series of four seminal papers \cite{BohPin-51,BohPin-52,BohPin-53,Pines-53} published in the early 1950s, Bohm and Pines proposed the random phase approximation (RPA) as an effective theory to describe the collective excitations of  jellium, a homogeneous high-density electron gas moving in a background
of uniform positive charge. In particular, they predicted that the electron gas will be decoupled into quasi-free electrons which emerge from the usual mean-field approximation for independent particles, and collective plasmon excitations which correspond to correlated particle motion.

\medskip

Although the plasmons were quickly detected by experiments \cite{Wat-56,Fer-57} after the works of Bohm and Pines, their theoretical explanation remains an important open question in condensed matter and nuclear physics. In 1957, Gell-Mann and Brueckner \cite{GelBru-57} gave a microscopic derivation of the RPA using a formal summation of a diagrammatic expansion, in which the leading diagrams describe the interaction of pairs of fermions, one from inside and one from outside the Fermi ball.
This approach was pushed further by Sawada \cite{Sawada-57} and Sawada--Brueckner--Fukuda--Brout
\cite{SawBruFukBro-57} who interpreted these pairs of electrons as bosons, obtaining an effective Hamiltonian which
is quadratic with respect to the bosonic particle pairs. 

\medskip

Recently, the bosonization argument in \cite{Sawada-57,SawBruFukBro-57} has been made rigorous in \cite{HaiPorRex-20,BNPSS-20,BNPSS-21,Benedikter-20,CHN-21,BPSS-22} for bounded interaction potentials in the mean-field regime, in which the interaction potential is coupled with a small constant such that the interaction energy and the kinetic energy are comparable. In these works, the non-bosonizable terms of the interaction energy are negligible and the rest can be diagonalized by adapting Bogolubov's method \cite{Bogolubov-47} to the quasi-bosonic setting. On the mathematical side, the main challenge in this approach is to realize the bosonization structure, which only holds in a very weak sense, making even perturbative results highly nontrivial  \cite{HaiPorRex-20}. In the first non-perturbative results in \cite{BNPSS-20,BNPSS-21}, the correlation energy was computed exactly to the leading order by using a patching technique (averaging fermionic pairs in patches of the Fermi sphere) to enhance the bosonization structure. This approach has been developed further in \cite{BPSS-22} to improve the analysis of the ground state energy and in \cite{BNPSS-22} to address the dynamics. In \cite{CHN-21} we proposed an alternative approach where the weak bosonization structure was used directly (without relying on the patching technique) to approximately diagonalize the fermionic Hamiltonian.  One of the advantages of this approach is that it allows us to derive an effective quasi-bosonic Hamiltonian which describes both the correlation energy and the elementary excitations of the system. In the mean-field regime there are however no approximate eigenstates corresponding to collective plasmon modes.

\medskip

The aim of the present paper is to give an explanation of the collective plasmon excitations by taking the quasi-bosonic Hamiltonian derived in \cite{CHN-21}, extrapolating for the Coulomb potential and going beyond the mean-field regime. We hope that our analysis here will provide useful insights towards the ultimate goal of deriving this effective Hamiltonian and understanding the plasmons from first principles.

\section{Derivation of the Effective Hamiltonian}

In this section we give a heuristic derivation of the effective quasi-bosonic Hamiltonian from the microscopic theory, by summarizing the approach in \cite{CHN-21}.

\medskip

We consider a system of $N$ (spinless) fermions on the torus $\mathbb{T}^{3}=\left[0,2\pi\right]^{3}$ (with periodic boundary conditions), interacting via a repulsive potential $V:\mathbb{T}^{3}\rightarrow\mathbb{R}$, which is to say 
\begin{equation}
V\left(x\right)=\frac{1}{\left(2\pi\right)^{3}}\sum_{k\in \mathbb{Z}^{3}_* }\hat{V}_{k}e^{ik\cdot x}\quad\text{with}\quad\hat{V}_{k}=\int_{\mathbb{T}^{3}}V\left(x\right)e^{-ik\cdot x}\,dx \ge 0, \quad \forall k\in  \mathbb{Z}^{3}_* = \mathbb{Z}^{3} \backslash\{0\},
\end{equation}
and which satisfies the square summability condition on the Fourier transform 
\begin{equation}
\sum_{k\in \mathbb{Z}^{3}_*} \hat V_k^2 <\infty.
\end{equation}
Here we ignore the contribution of the zero-momentum mode (equivalently we set $\hat V_0=0$) as it corresponds to a trivial energy shift of the system (physically, this is understood to be compensated for by the uniformly charged background). The reader may keep in mind the typical situation of the Coulomb potential where $\hat{V}_{k}=g\left|k\right|^{-2}$ with $g>0$, although our analysis applies to a larger class of potentials. 

\medskip

In the many-body Schr\"odinger theory, the system is described by the Hamiltonian 
\begin{equation} \label{eq:intro-HN}
H_{N}=H_{\text{kin}}+ H_{\text{int}} = \sum_{i=1}^{N} (-\Delta_{i}) + \sum_{1\leq i<j\leq N}V\left(x_{i}-x_{j}\right)
\end{equation}
which acts on the fermionic space
\begin{equation} \label{eq:cHN-space}
\mathcal{H}_{N}=\bigwedge^{N}\mathfrak{h}, \quad \mathfrak{h}=L^{2}\left(\mathbb{T}^{3}\right). 
\end{equation}
Under our assumption, $H_N$ is bounded from below and it can be extended to be a self-adjoint operator on $\mathcal{H}_{N}$ with domain 
$D\left(H_{N}\right)=D\left(H_{\text{kin}}\right)=\bigwedge^{N}H^{2}\left(\mathbb{T}^{3}\right).$ Moreover, $H_N$ has compact resolvent and we are interested in the  low-lying spectrum of $H_N$ when $N\to \infty$. 

\medskip
In general, if $V\not \equiv 0$ and $N$ is large, computing the spectrum of $H_N$ directly from the microscopic formulation \eqref{eq:intro-HN} is impossible, both analytically and numerically. Consequently one must turn to efficient approximations. One of the most famous approximations for fermions is Hartree--Fock
theory, where one restricts the consideration to Slater determinants $g_{1}\wedge g_{2}\cdots\wedge g_{N}$
with $\left\{ g_{i}\right\} _{i=1}^{N}$ orthonormal in $L^{2}\left(\mathbb{T}^{3}\right)$, which are the least correlated states among all
fermionic wave functions. The precision of the Hartree--Fock energy for Coulomb systems can be estimated using general {\em correlation inequalities} of Bach \cite{Bach-92} and Graf--Solovej \cite{GraSol-94}. Within Hartree--Fock theory, it turns out that the ground state energy can be well approximated by the Fermi state, which is the Slater determinant of the plane waves with momenta inside the Fermi ball $B_F$, namely
\begin{equation} \label{eq:intro-FS}
\psi_{{\rm FS}}=\bigwedge_{p\in B_{F}}u_{p}, \quad u_{p}\left(x\right)=\left(2\pi\right)^{-\frac{3}{2}}e^{ip\cdot x},
\end{equation}
with 
\begin{equation} \label{eq:intro-N}
B_{F}=\overline{B}\left(0,k_{F}\right)\cap\mathbb{Z}^{3}, \quad N=\left|B_{F}\right|,
\end{equation}
for some $k_F > 0$ (the Fermi momentum); see \cite{GonHaiLew-19} and \cite[Appendix A]{BNPSS-21}. Here for simplicity we assume that the Fermi ball $B_F$ is completely filled by $N$ integer points, 
which implies that the Fermi state $\psi_{{\rm FS}}$  is the unique, non-degenerate ground state of the kinetic operator $H_{\text{kin}}$. Without this simplification, the Fermi state is not uniquely defined and the degeneracy of the elementary excitation introduced in the next subsection has to be factored out properly, which complicate  the notation but do not improve the physical insight that we want to discuss. 
\medskip

In order to focus on the correlation
structure of the interacting system, we need to extract the energy of the Fermi state. For this purpose, it is convenient to write the second-quantized form of the Hamiltonian operator $H_{N}$ in \eqref{eq:intro-HN}: 
\begin{equation}
H_{N}=H_{\text{kin}}+H_{\text{int}} = \sum_{p\in\mathbb{Z}^{3}_*}\left|p\right|^{2}c_{p}^{\ast}c_{p}+\frac{1}{2\left(2\pi\right)^{3}}\sum_{k\in\mathbb{Z}^{3}_*}\sum_{p,q\in\mathbb{Z}^{3}}\hat{V}_{k}c_{p+k}^{\ast}c_{q-k}^{\ast}c_{q}c_{p}\label{eq:IntroductionSecondQuantizedHamiltonian}
\end{equation}
where
\begin{equation}
c_{p}^{\ast}=c^*(u_p), \quad c_{p}=c(u_p), \quad \forall p\in \mathbb{Z}^3,
\end{equation}
are the usual Fermionic creation and annihilation operators associated to the plane wave states $u_p$. Note that although the second-quantized form in \eqref{eq:IntroductionSecondQuantizedHamiltonian} can be defined on the fermionic Fock space, we will always consider its restriction to the $N$ particle space which coincides with the original Hamiltonian in \eqref{eq:intro-HN}. 

Using the canonical anticommutation relations (CAR)
\begin{equation} \label{eq:CAR}
\left\{ c_{p},c_{q}\right\} =\left\{ c_{p}^{\ast},c_{q}^{\ast}\right\} =0,\quad\left\{ c_{p},c_{q}^{\ast}\right\} =\delta_{p,q},\quad p,q\in\mathbb{Z}^{3},
\end{equation}
where $\left\{ A,B\right\} =AB+BA$, it is straightforward to compute the energy of the Fermi state (see e.g. \cite[Eq. (1.10) and Eq. (1.20)]{CHN-21})
\begin{align}
E_{\rm FS} = \langle \psi_{{\rm FS}}, H_{N} \psi_{{\rm FS}} \rangle &=   \left\langle \psi_{{\rm FS}},H_{\text{kin}}\psi_{{\rm FS}}\right\rangle + \left\langle \psi_{{\rm FS}},H_{\text{int}}\psi_{{\rm FS}}\right\rangle \nonumber\\
&=\sum_{p\in B_{F}}\left|p\right|^{2} + \frac{1}{2\left(2\pi\right)^{3}}\sum_{k\in\mathbb{Z}_{\ast}^{3}}\hat{V}_{k}\left(\left|L_{k}\right|-N\right)
\end{align}
where we define the {\em lune} of relative momentum $k\in \mathbb{Z}^3_*$ by  
\begin{equation} \label{eq:lune}
L_k = B_{F}^{c}\cap\left(B_{F}+k\right)= \left\{ p\in\mathbb{Z}^{3}\mid\left|p-k\right|\leq k_{F}<\left|p\right|\right\}.
\end{equation}

\bigskip

Now we extract the contribution of the Fermi state on the operator level, namely we rewrite the operator in \eqref{eq:IntroductionSecondQuantizedHamiltonian} as 
\begin{align} \label{eq:HN-localized-1}
H_{N}  = E_{\rm FS} +H_{\kin}^{\prime}+ H_{\inter}^{\prime}
\end{align}
for suitable operators $H_{\kin}^{\prime}, H_{\inter}^{\prime}:D\left(H_{\text{kin}}\right)\subset\mathcal{H}_{N}\rightarrow\mathcal{H}_{N}$. To be precise, we define the {\em localized kinetic operator}  as 
\begin{align} \label{eq:Hkin'}
H_{\kin}^{\prime}&=H_{\text{kin}}  - \left\langle \psi_{{\rm FS}},H_{\text{kin}}\psi_{{\rm FS}}\right\rangle \ge 0 
\end{align}
and define the {\em localized interaction operator}  as 
\begin{align}\label{eq:localizedInteractionBandDForm}
H_{\text{int}}^{\prime} &=H_{\text{int}}- \left\langle \psi_{{\rm FS}},H_{\text{int}}\psi_{{\rm FS}}\right\rangle = 
\sum_{k\in\mathbb{Z}_{*}^{3}}\left(H_{\text{int}}^{k}-\frac{\hat{V}_{k}}{2\left(2\pi\right)^{3}}\left|L_{k}\right|\right)\nonumber\\
&\quad  +\frac{1}{\left(2\pi\right)^{3}}\sum_{k\in\mathbb{Z}_{\ast}^{3}}\hat{V}_{k}\left( \sum_{p\in L_{k}} b_{k,p}^* D_{k}+D_{k}^{\ast}\sum_{p\in L_{k}} b_{k,p}+\frac{1}{2}D_{k}^{\ast}D_{k}\right)  
\end{align}
where 
\begin{align} \label{eq:Hint-k}
H_{\text{int}}^{k} =\sum_{p,q\in L_{k}}\frac{\hat{V}_{k}}{2\left(2\pi\right)^{3}}\left(b_{k,p}^{\ast}b_{k,q}+b_{k,q}b_{k,p}^{\ast}\right)+\sum_{p\in L_{k}}\sum_{q\in L_{-k}}\frac{\hat{V}_{k}}{2\left(2\pi\right)^{3}}\left(b_{k,p}^{\ast}b_{-k,q}^{\ast}+b_{-k,q}b_{k,p}\right)
\end{align}
for
\begin{align}  \label{eq:def-bkp}
b_{k,p}^*=c_{p}^{\ast}c_{p-k}, \quad D_k =  \sum_{p\in  B_F \cap (B_F+k)} c_{p-k}^{\ast}c_{p} +  \sum_{p\in B_F^c \cap (B_F^c+k) } c_{p-k}^{\ast}c_{p}. 
\end{align}
We interpret $b_{k,p}^*$ as an  \textit{excitation operator}, since it creates a state
with momentum $p\in B_{F}^{c}$ and annihilates a state with momentum $p-k\in B_{F}$.

\subsection{The Effective Quasi-Bosonic Hamiltonian}

So far, the decomposition of \eqref{eq:HN-localized-1} is exact, but to proceed further we now make some simplifications. Roughly speaking, the RPA in the physics literature \cite{GelBru-57,Sawada-57,SawBruFukBro-57} suggests that the fermionic correlation structure can be described by a bosonic quadratic Hamiltonian. As explained in \cite{CHN-21}, this bosonic analogy can be summarized in three steps:

\textbf{Step 1.} The excitation operators $b_{k,p}^{\ast}$, $b_{k,p}$ in \eqref{eq:def-bkp} 
should be treated as \textit{bosonic} creation and annihilation operators,
where the operators $b_{k,p}$ and $b_{l,q}$ with $k\neq l$ can
be considered as acting on independent Fock spaces.

On the mathematical side, we expect the canonical commutation relations
(CCR) to hold in an appropriate sense: 
\begin{equation}
\left[b_{k,p},b_{l,q}\right]=\left[b_{k,p}^{\ast},b_{l,q}^{\ast}\right]=0,\quad\left[b_{k,p},b_{l,q}^{\ast}\right]\approx\delta_{k,l}\delta_{p,q}.\label{eq:IntroductionApproximateCCR}
\end{equation}

To motivate \eqref{eq:IntroductionApproximateCCR}, let us consider the simple case $k=l$ where we have the exact relations
\begin{equation} \label{eq:intro-comm-b-b*}
\left[b_{k,p},b_{k,q}\right]=[b_{k,p}^{\ast},b_{k,q}^{\ast}]=0, \quad \left[b_{k,p},b_{k,q}^{\ast}\right]=\delta_{p,q}-\delta_{p,q}\left(c_{p}^{\ast}c_{p}+c_{p-k}c_{p-k}^{\ast}\right)
\end{equation}
for all $p,q\in L_{k}$.  The last error terms in \eqref{eq:intro-comm-b-b*} are not small individually (as we only know $c_{p}^{\ast}c_{p},\,c_{p}c_{p}^{\ast}\leq1$ by Pauli's exclusion principle), but they are small on average. To make it transparent, let us introduce the {\em excitation number operator} 
\begin{equation}
\mathcal{N}_{E}:=\sum_{p\in B_{F}^{c}}c_{p}^{\ast}c_{p}=\sum_{p\in B_{F}}c_{p}c_{p}^{\ast}\quad\text{on }\mathcal{H}_{N}\label{eq:ParticleHoleSymmetry}
\end{equation}
where the last identity in \eqref{eq:ParticleHoleSymmetry} follows from the assumption $|B_F|=N$ via the \textit{particle-hole symmetry}\footnote{Namely, the \textit{excitation number operator} (which counts the number of particles outside the Fermi state) coincides with the \textit{hole number operator} (which counts the number of holes inside the Fermi state).}. Then it is obvious that 
\begin{equation}
\sum_{p,q\in L_{k}}\delta_{p,q}\left(c_{p}^{\ast}c_{p}+c_{p-k}c_{p-k}^{\ast}\right) \leq2\,\mathcal{N}_{E}
\end{equation}
while for the low-lying eigenfunctions of $H_N$ the excitation number operator  $\mathcal{N}_{E}$ is expected to be of lower order than
\begin{align} \label{eq:|Lk|}
\sum_{p,q\in L_{k}}\delta_{p,q}=\left|L_{k}\right| \sim \min (|k|,k_F)k_F^2.
\end{align}
See e.g. \cite[Proposition A.1]{CHN-21} for estimates related to \eqref{eq:|Lk|}. 

\medskip

\textbf{Step 2.} The full operator in (\ref{eq:HN-localized-1}) is approximated by a quadratic Hamiltonian of $b_{k,p}^{\ast}$ and $b_{k,p}$. Concretely, the \textit{non-bosonizable terms}, which are the last sum in \eqref{eq:localizedInteractionBandDForm}, are ignored, so that
\begin{align} 
H_{\text{int}}^{\prime} \approx 
\sum_{k\in\mathbb{Z}_{*}^{3}}\left(H_{\text{int}}^{k}-\frac{\hat{V}_{k}}{2\left(2\pi\right)^{3}}\left|L_{k}\right|\right)
\end{align}
with $H_{\text{int}}^{k}$ given in \eqref{eq:Hint-k},  and the localized kinetic operator is thought of as
\begin{equation} \label{eq:intro-kinetic-bosonic}
H_{\kin}^{\prime}\approx \sum_{k\in\mathbb{Z}_{\ast}^{3}}\sum_{p\in L_{k}} 2 \lambda_{k,p}  b_{k,p}^{\ast}b_{k,p}, \quad  \lambda_{k,p}=\frac{1}{2}(\left|p\right|^{2}-\left|p-k\right|^{2}).
\end{equation}
The latter approximation \eqref{eq:intro-kinetic-bosonic} is motivated by the commutation relations
\begin{equation} \label{eq:intro-Hkin-com}
\left[H_{\kin}^{\prime},b_{k,p}^{\ast}\right]= 2 \lambda_{k,p} b_{k,p}^{\ast} \approx \left[ \sum_{\ell \in\mathbb{Z}_{\ast}^{3}}\sum_{q\in L_{\ell}} 2 \lambda_{\ell,q}  b_{\ell,q}^{\ast}b_{\ell,q} , b_{k,p}^{\ast}\right] 
\end{equation}
where the first identity follows from the (exact) CAR \eqref{eq:CAR} and the second relation follows from the (approximate) CCR \eqref{eq:IntroductionApproximateCCR}.

\medskip

\textbf{Step 3.} If the effective Hamiltonian 
\begin{align} \label{eq:intro-eff-Hamil-quasi-bosonic}
\sum_{k\in\mathbb{Z}_{\ast}^{3}}\sum_{p\in L_{k}} 2 \lambda_{k,p}  b_{k,p}^{\ast}b_{k,p} + 
\sum_{k\in\mathbb{Z}_{*}^{3}}\left(H_{\text{int}}^{k}-\frac{\hat{V}_{k}}{2\left(2\pi\right)^{3}}\left|L_{k}\right|\right)
\end{align}
were an exact bosonic quadratic operator, then it could be diagonalized by a Bogolubov transformation (see e.g. \cite[Section 3.2]{CHN-21}), resulting in the effective operator  
\begin{align} \label{eq:effective}
E_{\rm corr} + \sum_{k\in\mathbb{Z}_{\ast}^{3}}  2 \sum_{p,q\in L_{k}}\left\langle e_{p}, \tilde{E}_{k} e_{q}\right\rangle b_{k,p}^{\ast}b_{k,q}.  
\end{align}
Here we introduced the correlation energy 
\begin{equation} \label{eq:correlation-energy}
E_{\rm corr}  = \sum_{k\in \mathbb{Z}^3_*} \left(  \text{tr}\left(\widetilde{E}_{k}-h_{k}\right)-\frac{\hat{V}_{k}}{2\left(2\pi\right)^{3}}\left|L_{k}\right| \right)=\sum_{k\in \mathbb{Z}^3_*}  \frac{1}{\pi}\int_{0}^{\infty}F\left(\frac{\hat{V}_{k}}{\left(2\pi\right)^{3}}\sum_{p\in L_{k}}\frac{\lambda_{k,p}}{\lambda_{k,p}^{2}+t^{2}}\right)dt
\end{equation}
with $F\left(x\right)=\log\left(1+x\right)-x$, and  for every $k\in\mathbb{Z}_{\ast}^{3}$ we defined the following real, symmetric operators on $\ell^{2}\left(L_{k}\right)$: 
\begin{equation}
\widetilde{E}_{k}= (h_{k}^{\frac{1}{2}}\left(h_{k}+2P_{v_{k}}\right)h_{k}^{\frac{1}{2}})^{\frac{1}{2}}, \quad h_{k}e_{p}=\lambda_{k,p}e_{p}, \quad  P_{v}  =|v_{k} \rangle \langle v_k|,  \quad v_{k}=\sqrt{\frac{\hat{V}_{k}}{2\left(2\pi\right)^{3}}}\sum_{p\in L_{k}}e_{p},\label{eq:IntroductionOneBodyOperators}
\end{equation}
with $\left(e_{p}\right)_{p\in L_{k}}$ the standard orthonormal basis of $\ell^{2}\left(L_{k}\right)$. However, 
the quadratic kinetic approximation of \eqref{eq:intro-kinetic-bosonic} only holds in the weak sense of  \eqref{eq:intro-Hkin-com}, so the difference 
\begin{equation}  \label{eq:diff-Hkin-Hbo}
H_{\kin}^{\prime}- \sum_{k\in\mathbb{Z}_{\ast}^{3}}\sum_{p\in L_{k}} 2 \lambda_{k,p}  b_{k,p}^{\ast}b_{k,p} 
\end{equation}
is only essentially invariant under the Bogolubov transformation, rather than close to $0$ in a direct sense. Therefore, adding \eqref{eq:diff-Hkin-Hbo} to \eqref{eq:effective} we obtain the more realistic approximation,  up to a unitary transformation,  that
\begin{align} \label{eq:IntroductionFullRPA}
 H_{N}  \approx E_{\rm FS}  + E_{\rm corr} + H_{\rm eff},  
\end{align}
where we introduced the effective quasi-bosonic Hamiltonian 
\begin{equation} \label{eq:Heff}
H_{\rm eff}= H_{\kin}^{\prime}  + 2 \sum_{k\in\mathbb{Z}_{\ast}^{3}}  \sum_{p,q\in L_{k}}\left\langle e_{p}, (\tilde{E}_{k} -h_k) e_{q}\right\rangle b_{k,p}^{\ast}b_{k,q}. 
\end{equation}
which is an operator on the fermionic space $\mathcal{H}_{N}=\bigwedge^{N}\mathfrak{h}$. 

\medskip

All in all the bosonization procedure of the random phase approximation thus suggests that \eqref{eq:IntroductionFullRPA} holds at least for states with few excitations (when $\mathcal{N}_E$ is not too large).

\medskip

For regular potentials in the mean-field regime, i.e. when $V$ is replaced by $k_F^{-1} W$ for a fixed potential $W$ satisfying $\sum_{k\in \mathbb{Z}^3_*} |k| |\hat W(k)| <\infty$, the operator approximation \eqref{eq:IntroductionFullRPA} has been justified rigorously in \cite{CHN-21}. To be precise, we proved in \cite[Theorem 1] {CHN-21} that there exists a unitary operator $\mathcal{U}: \mathcal{H}_N\to \mathcal{H}_N$ such that
\begin{align} \label{eq:FullRPA-ope}
 \mathcal{U} H_{N} \mathcal{U}^* = E_{\rm FS}  + E_{\rm corr} + H_{\rm eff} +\mathcal{E}_{\mathcal{U}}
\end{align}
where the error operator satisfies
\begin{align} \label{eq:FullRPA-err}
\pm \mathcal{E}_{\mathcal{U}}  \le C k_F^{-\frac 1 {94} + \epsilon} (k_F^{-1} \mathcal{N}_E H_{\rm kin}'+ H_{\rm kin}' + k_F), \quad k_F\to \infty,
\end{align}
for any fixed $\epsilon >0$. Moreover, thanks to \cite[Theorem 1.2]{CHN-21}, the bound in \eqref{eq:FullRPA-err} suffices to show that $\mathcal{E}_{\mathcal{U}}$ is negligible when applied to low-lying eigenstates $\Psi$ of $\mathcal{U} H_{N} \mathcal{U}^*$ satisfying $\langle \Psi, \mathcal{U} H_{N} \mathcal{U}^* \Psi\rangle = E_{\rm FS}  + E_{\rm corr} + O(k_F)$, namely 
\begin{align} \label{eq:FullRPA-err-2}
| \langle \Psi, \mathcal{E}_{\mathcal{U}}  \Psi\rangle| \le C k_F^{1-\frac 1 {94} + \epsilon}, \quad k_F\to \infty,
\end{align}
while both $E_{\rm corr}$ and $\langle \Psi, H_{\rm eff} \Psi\rangle$ are of order $k_F$.  

\medskip 

Note that even in the mean-field regime, the Coulomb potential is still excluded in \cite{CHN-21}. In this case, when $\hat V_k$ is replaced by $g k_F^{-1} |k|^{-2}$, the correlation energy $E_{\rm corr}$ is of order $k_F \log(k_F)$ instead of $k_F$, and existing techniques seem insufficient to estimate the error terms for the energy lower bound. We refer to the recent work \cite{CHN-22} for a rigorous upper bound for the correlation energy. The operator approximation \eqref{eq:IntroductionFullRPA} for the Coulomb gas in the mean-field regime remains completely open, let alone the corresponding result beyond the mean-field regime. 

\medskip 

In the present paper, we will consider the effective operator $H_{\rm eff}$ in more detail, without imposing the mean-field and regularity restrictions on the interaction. In particular, we will focus on the most interesting case of the Coulomb potential $\hat V_k = g|k|^{-2}$ for which the plasmon modes can be interpreted as a special class of approximate eigenstates of $H_{\rm eff}$.

\subsection{Elementary Excitations and the Plasmon Frequency}

As explained in  \cite{CHN-21}, since  the effective Hamiltonian $H_{\rm eff}$ in \eqref{eq:Heff} commutes with $\mathcal{N}_{E}$,  we can without loss of generality restrict $H_{\text{eff}}$ to the eigenspaces $\{\mathcal{N}_{E}=M\}$ with $M=0,1,2,...$

The case $M=0$ is trivial since the eigenspace $\{\mathcal{N}_E=0\}$ is the one-dimensional space spanned by the Fermi state. In the first non-trivial case, $M=1$, the identity
\begin{align}
 \sum_{k\in\mathbb{Z}_{\ast}^{3}}\sum_{p\in L_{k}} 2 \lambda_{k,p}  b_{k,p}^{\ast}b_{k,p}  = \mathcal{N}_{E} H_{\kin}^{\prime}
\end{align}
(see \cite[Eq. (1.55)]{CHN-21}) implies that the relation of  \eqref{eq:intro-kinetic-bosonic} is in fact valid, whence
\begin{equation}
\left.H_{\text{eff}}\right|_{\mathcal{N}_{E}=1}=2\sum_{k\in\mathbb{Z}_{\ast}^{3}}\sum_{p,q\in L_{k}}\left\langle e_{p},\widetilde{E}_{k}e_{q}\right\rangle b_{k,p}^{\ast}b_{k,q}.
\end{equation}
This operator can be diagonalized explicitly on $\{\mathcal{N}_{E}=1\}$. More precisely, it was proved in \cite[Theorem 1.4]{CHN-21} that by introducing the unitary transformation 
\begin{equation} 
\tilde{U}:\bigoplus_{k\in\mathbb{Z}_{\ast}^{3}}\ell^{2}\left(L_{k}\right)\rightarrow\left\{ \Psi\in\mathcal{H}_{N}\mid\mathcal{N}_{E}\Psi=\Psi\right\},  
\end{equation}
\begin{equation} \label{eq:intro-tilde-U}
\tilde{U}\bigoplus_{k\in\mathbb{Z}_{\ast}^{3}}\varphi_{k}=\sum_{k\in\mathbb{Z}_{\ast}^{3}}b_{k}^{\ast}\left(\varphi_{k}\right)\psi_{\rm FS},
\end{equation}
where for any $\varphi \in \ell^2(L_k)$ the \textit{generalized excitation operator} $b_k^\ast(\varphi)$ is defined by
\begin{equation} \label{eq:GEOdef}
b_k^\ast(\varphi) = \sum_{p \in L_k} \left\langle e_p , \varphi \right\rangle b_{k,p}^\ast ,
\end{equation}
we have the identity 
 \begin{equation} \label{eq:intro-excitation-NE1}
\tilde{U}^{\ast} \left( \left.H_{\rm eff}\right|_{\mathcal{N}_{E}=1} \right) \tilde{U} =   \bigoplus_{k\in\mathbb{Z}_{\ast}^{3}}2\widetilde{E}_{k}  \quad \text{ on }\bigoplus_{k\in\mathbb{Z}_{\ast}^{3}}\ell^{2}\left(L_{k}\right). 
\end{equation}
Consequently, the spectrum of  $\left.H_{\rm eff}\right|_{\mathcal{N}_{E}=1}$ is fully determined by the eigenvalues of $2\widetilde{E}_{k}$. Note that every eigenvalue $\epsilon$ of  $2\widetilde{E}_{k}=2(h_{k}^{\frac{1}{2}}\left(h_{k}+2P_{v_{k}}\right)h_{k}^{\frac{1}{2}})^{\frac{1}{2}}$ solves the equation 
\begin{equation}
(\epsilon^{2} - 4h_{k}^{2})  w=4\widetilde{E}_{k}^2 w- 4h_{k}^{2}w = 8\langle h_{k}^{\frac{1}{2}}v_{k},w\rangle h_{k}^{\frac{1}{2}}v_{k}
\end{equation}
for a normalized eigenvector $w$. Therefore, if $\epsilon$ is not an eigenvalue of $2h_{k}$, we can take the inner product with $\langle h_{k}^{\frac{1}{2}}v_{k},w\rangle^{-1}(\epsilon^{2} - 4 h_{k}^{2})^{-1}h_{k}^{\frac{1}{2}}v_{k}$ and obtain 
\begin{align} \label{eq:ev-Ek}
1=8\langle v_{k},h_{k}\left(\epsilon^{2}- 4 h_{k}^{2}\right)^{-1}v_{k}\rangle&=\frac{4\hat{V}_{k}}{\left(2\pi\right)^{3}}\sum_{p\in L_{k}}\frac{\lambda_{k,p}}{\epsilon^{2}- 4 \lambda_{k,p}^{2}} \nonumber\\
& = \frac{2\hat{V}_{k}}{\left(2\pi\right)^{3}}\sum_{p\in B_{F}}\frac{\left|k\right|^{2}}{\left(\epsilon-2 k\cdot p\right)^{2}-\left|k\right|^{4}},
\end{align}
which coincides with \cite[Eq. (6)]{SawBruFukBro-57}. Since the last equality in \eqref{eq:ev-Ek} is not obvious, let us add an explanation for the reader's convenience. Using the algebraic identity
\begin{equation}
\frac{\lambda_{k,p}}{\epsilon^{2}- 4 \lambda_{k,p}^{2}}  = \frac{\lambda_{k,p}}{(\epsilon- 2 \lambda_{k,p}) (\epsilon+ 2 \lambda_{k,p}) }  = \frac{1}{4} \left( \frac{1}{\epsilon- 2 \lambda_{k,p}} - \frac{1}{ \epsilon+ 2 \lambda_{k,p}} \right)   
\end{equation}
and the definition $L_k = B_F^c \cap (B_F+k) = (B_F + k) \backslash B_F$ we can write
\begin{align} \label{eq:proof-SBFB-e}
&\sum_{p\in L_{k}}\frac{\lambda_{k,p}}{\epsilon^{2}- 4 \lambda_{k,p}^{2}} = \frac{1}{4}\sum_{p\in L_{k}} \left( \frac{1}{\epsilon- 2 \lambda_{k,p}} - \frac{1}{ \epsilon+ 2 \lambda_{k,p}} \right) \\
&  = \frac{1}{4}\sum_{p\in (B_F+k)} \left( \frac{1}{\epsilon- 2 \lambda_{k,p}} - \frac{1}{ \epsilon+ 2 \lambda_{k,p}} \right) - \frac{1}{4}\sum_{p\in (B_F+k)\cap B_F} \left( \frac{1}{\epsilon- 2 \lambda_{k,p}} - \frac{1}{ \epsilon+ 2 \lambda_{k,p}} \right)\nonumber.
\end{align}
Using $2\lambda_{k,p}=2k\cdot ( p - \frac k 2)$ and substituting $p\mapsto p-\frac{1}{2}k$ in the last sum in \eqref{eq:proof-SBFB-e} we get 
\begin{align}
&\frac{1}{4} \sum_{p\in (B_F+k)\cap B_F} \left( \frac{1}{\epsilon- 2 \lambda_{k,p}} - \frac{1}{ \epsilon+ 2 \lambda_{k,p}} \right) \nonumber\\
&= \frac{1}{4} \sum_{p\in (B_F+\frac k 2)\cap (B_F - \frac k 2) } \left( \frac{1}{\epsilon- 2 k \cdot p} - \frac{1}{ \epsilon+ 2 k \cdot p} \right)=0
\end{align}
where the cancelation comes from the symmetry $p\mapsto -p$. On the other hand, by substituting $p\mapsto p+k$ in the first sum in \eqref{eq:proof-SBFB-e} we can write
\begin{align}
&\frac{1}{4}\sum_{p\in (B_F+k)} \left( \frac{1}{\epsilon- 2 \lambda_{k,p}} - \frac{1}{ \epsilon+ 2 \lambda_{k,p}} \right) = \frac{1}{4} \sum_{p\in B_F} \left( \frac{1}{\epsilon- 2 k \cdot p - |k|^2} - \frac{1}{ \epsilon+ 2 k \cdot p + |k|^2} \right)\nonumber\\
&=\frac{1}{4}\sum_{p\in B_F} \left( \frac{1}{\epsilon- 2 k \cdot p - |k|^2} - \frac{1}{ \epsilon - 2 k \cdot p + |k|^2} \right)=\frac{1}{2}\sum_{p\in B_F}   \frac{|k|^2}{ (\epsilon- 2 k \cdot p)^2 - |k|^4} 
\end{align}
where we also transformed $p\mapsto -p$ on the second term. Thus \eqref{eq:ev-Ek} holds. 
\bigskip

In summary, \eqref{eq:ev-Ek} characterizes all eigenvalues of $2 \widetilde{E}_{k}$ outside the spectrum of $2 h_k$. In the case of the Coulomb potential $\hat{V}_{k} = g\left|k\right|^{-2}$,  with a constant $g>0$, the $k$-dependence in \eqref{eq:ev-Ek} is simplified and we obtain
\begin{align} \label{eq:ev-Ek-Coulomb}
1=\frac{2g}{\left(2\pi\right)^{3}}\sum_{p\in B_{F}}\frac{1}{\left(\epsilon-2 k\cdot p\right)^{2}-\left|k\right|^{4}}.
\end{align}
In this case, among all eigenvalues described in \eqref{eq:ev-Ek-Coulomb}, the largest one is special as it is proportional to $k_F^{3/2}$ while the other eigenvalues are bounded from above by 
\begin{equation} \label{eq:2.45}
2 \lambda_{k, \max}:=\sup_{p\in B_F} (2 k\cdot p + |k|^2) \le 2|k| k_F + |k|^2 \ll k_F^{3/2} \quad \text  {if} \quad |k|\ll k_F^{1/2}.
\end{equation}
Indeed, note that the function 
\begin{equation}
f(\epsilon) = \frac{2g}{\left(2\pi\right)^{3}}\sum_{p\in B_{F}}\frac{1}{\left(\epsilon-2 k\cdot p\right)^{2}-\left|k\right|^{4}}
\end{equation}
is strictly decreasing on $(2 \lambda_{k, \max},\infty)$ and 
\begin{equation}
\lim_{\epsilon\to (2 \lambda_{k, \max})^+ }f(\epsilon) =\infty, \quad  \lim_{\epsilon\to \infty}f(\epsilon) =0.
\end{equation} Therefore, the equation $f(\epsilon)=1$ has a unique solution on $(2 \lambda_{k, \max},\infty)$. Moreover, this solution satisfies
\begin{equation}
\epsilon^2 = \frac{2g}{\left(2\pi\right)^{3}}\sum_{p\in B_{F}}\frac{\epsilon^2}{\left(\epsilon-2 k\cdot p\right)^{2}-\left|k\right|^{4}} \ge \frac{2g}{\left(2\pi\right)^{3}}\sum_{p\in B_{F}}1 = \frac{2g N}{\left(2\pi\right)^{3}}
\end{equation}
namely
\begin{equation} \label{eq:eps-lower-bound-first}
\epsilon \ge \sqrt{\frac{2gN}{(2\pi)^3}} = \sqrt{\frac{g}{3\pi^2}} k_F^{3/2}  (1 + o(1)_{k_F\to \infty}). 
\end{equation}
When $|k|\ll k_F^{1/2}$, the lower bound in \eqref{eq:eps-lower-bound-first} implies that 
\begin{equation}
\frac{\epsilon^2}{\left(\epsilon-2 k\cdot p\right)^{2}-\left|k\right|^{4}} \approx 1,
\end{equation}
and hence \eqref{eq:eps-lower-bound-first} is asymptotically sharp, namely we have 
\begin{equation}
\epsilon = \sqrt{\frac{g}{3\pi^2}} k_F^{3/2}  (1 + o(1)_{k_F\to \infty}). 
\end{equation}
In summary, if $|k|\ll k_F^{1/2}$, then the largest eigenvalue $\epsilon$ of $2 \widetilde{E}_{k}$ is proportional to $k_F^{3/2}$, while all other eigenvalues of $2 \widetilde{E}_{k}$, either being characterized by \eqref{eq:ev-Ek-Coulomb} or belonging to the spectrum of $2 h_k$, are always bounded by $2 \lambda_{k, \max} \ll k_F^{3/2}$.  

\bigskip

In the physics literature, the largest eigenvalue of $2\widetilde{E}_{k}$ is often computed in the thermodynamic limit, where we replace Riemann sums by integrals and obtain 
\begin{align} \label{eq:plasmon-square}
\epsilon^{2} \approx 2gn
\end{align}
where 
\begin{equation}
n=\frac{N}{\mathcal{V}}= \frac{\text{Vol}\left({B}\left(0,k_{F}\right)\right)} {(2\pi)^3} = \frac{1}{6\pi^{2}}k_{F}^{3}
\end{equation}

is the number density of the system\footnote{Here we consider the spinless fermions for simplicity. If we include a factor of $q$ for the electron spin states (e.g. $q=2$ for electrons), the equality \eqref{eq:plasmon-square} is still correct provided that $n=\frac{q}{6\pi^{2}}k_{F}^{3}$.}. By taking $g=4\pi e^{2}$, and also inserting $\frac{\hbar^{2}}{2m}=1$, we find that the largest eigenvalue of $2\widetilde{E}_{k}$ is 
\begin{equation} \label{eq:plasmon}
\epsilon\approx \sqrt{2gn}=  \hbar \sqrt{\frac{4\pi ne^{2}}{m}} = \hbar  \omega_0
\end{equation}
where $\omega_0=\sqrt{\frac{4\pi ne^{2}}{m}}$ is exactly the plasmon frequency written in \cite[Eq. (3-90)]{Pines-99} and \cite[Eq.  (15.16) - (15.18)]{Fetter-Walecka-71}. 
\bigskip

In the present paper, we will study $H_{\rm eff}$ in \eqref{eq:Heff} for a general $M\ge 1$. In this case, the spectrum of $H_{\rm eff}$ corresponds to not only the {\em elementary excitations} but also all of the {\em collective excitations} of the system. Unlike the simple case $M=1$ discussed above, for $M\ge 2$  the operator $\left.H_{\text{eff}}\right|_{\mathcal{N}_{E}=M}$ can not be diagonalized explicitly as in \eqref{eq:intro-excitation-NE1}, and hence understanding the spectrum of $H_{\text{eff}}$ is both interesting and difficult. We will focus on the part of the spectrum of $\left.H_{\text{eff}}\right|_{\mathcal{N}_{E}=M}$ which can be interpreted as describing the collective plasmon modes.

\section{Main results}

Consider the effective Hamiltonian $H_{\rm eff}$ in \eqref{eq:Heff}, i.e.
\begin{equation}
H_{\rm eff}= H_{\kin}^{\prime}  + 2 \sum_{k\in\mathbb{Z}_{\ast}^{3}}  \sum_{p,q\in L_{k}}\left\langle e_{p}, (\tilde{E}_{k} -h_k) e_{q}\right\rangle b_{k,p}^{\ast}b_{k,q},
\end{equation}
which is an operator on the fermionic $N$-particle space $\mathcal{H}_N= \bigwedge^{N}L^{2}\left(\mathbb{T}^{3}\right)$ with domain $D(H_N)=\bigwedge^{N}H^{2}\left(\mathbb{T}^{3}\right)$.

As discussed above, for $M=1$ the eigenfunctions of $H_\mathrm{eff}$ are precisely the states of the form $b_k^\ast(\varphi) \psi_{\FS}$, where $\varphi \in \ell^2(L_k)$ is an eigenvector of $2\widetilde{E}_k$. In the exact bosonic case, the eigenfunctions of $\left.H_{\rm{eff}}\right|_{\mathcal{N}_{E}=M}$ would be the states of the form
\begin{equation}
b_{k_1}^\ast(\varphi_1) \cdots b_{k_M}^\ast(\varphi_M) \psi_{\FS} 
\end{equation}
where each $\varphi_i \in \ell^2(L_k)$ is an eigenvector of $2 \widetilde{E}_{k_i}$ for $1 \leq i \leq M$.

For the effective Hamiltonian this is generally no longer true when $M \geq 2$. However, we will show that in the case that $k_1 = \cdots = k_M =: k$, where $\varphi_1 = \cdots = \varphi_M =: \varphi_k$ is the eigenvector of the greatest eigenvalue of $2 \widetilde{E}_k$ (and so describes the plasmon mode), this is nonetheless approximately correct.

For the specific case of the Coulomb potential, we prove the following:

\begin{thm}
\label{thm:CoulombPlasmonStates} Let $k_F>0$ be a large parameter. Let $\hat{V}_{k}=g\left|k\right|^{-2}$ with a constant $g>0$. Let $\delta\in\left(0,\frac{1}{2}\right)$, $\varepsilon\in\left(0,1\right)$, $\left|k\right|\leq k_{F}^{\delta}$ and $1\le M\leq k_{F}^{\varepsilon}$ be given. Let $\varphi_k \in \ell^2(L_k)$ denote the normalized eigenvector corresponding to the greatest eigenvalue, $\epsilon_k$, of $2 \widetilde{E}_k$, and define $\Psi_M \in \left\{ \mathcal{N}_{E}=M\right\}$ by 
$$\Psi_M = b_k^\ast(\varphi_k)^M \psi_{\FS}.$$
Then the normalized state $\hat{\Psi}_M = \Vert {\Psi}_M \Vert^{-1} {\Psi}_M$ obeys
\[
\left\Vert \left(H_{\mathrm{eff}}-M\epsilon_{k}\right)\hat{\Psi}_M\right\Vert \leq C\left|k\right|^{-1}\sqrt{k_{F}}M^{\frac{5}{2}}
\]
for a constant $C > 0$ depending only on $\delta$ and $\varepsilon$. Furthermore, it holds that 
\[
\epsilon_{k} = \sqrt{8\left\langle v_{k},h_{k}v_{k}\right\rangle +4\frac{\left\langle v_{k},h_{k}^{3}v_{k}\right\rangle }{\left\langle v_{k},h_{k}v_{k}\right\rangle }}+ O(k_{F}^{-\frac{1}{2}}\left|k\right|^{4}).
\]
\end{thm}

Here $O(k_{F}^{-\frac{1}{2}}\left|k\right|^{4})$ is a quantity that is bounded in absolute value by $k_{F}^{-\frac{1}{2}}\left|k\right|^{4}$ times a constant independent of $k_F$ and $k$. 

This theorem shows that we can consider $\hat{\Psi}_M$ to be an ``approximate eigenfunction'' of $H_\mathrm{eff}$ with ``approximate eigenvalue'' $M \epsilon_k$, when $M$ is not too large. Let us give some quick remarks on this theorem: 

\bigskip

{\bf 1.} The norm estimate implies both a dynamic and a spectral estimate: Owing to the elementary time evolution estimate $\Vert (e^{-itH} - e^{-itE}) \psi \Vert \leq \Vert (H - E) \psi \Vert t$ this shows that
\begin{equation}
\Vert (e^{-itH_\mathrm{eff}} - e^{-itM\epsilon_k}) \hat{\Psi}_M \Vert \ll 1 \quad \text{for} \quad M \epsilon_k t \ll C \frac{M \epsilon_k}{\vert k \vert^{-1} \sqrt{k_F} M^\frac{5}{2}} \sim M^{-\frac{3}{2}} k_F \vert k \vert;
\end{equation}
note that $(M \epsilon_k)^{-1}$ is the characteristic timescale of the oscillation of $\hat{\Psi}_M$, so this is a non-trivial statement for $M \ll (k_F \vert k \vert)^\frac{2}{3}$.

Spectrally, thanks to the operator inequality $|{\mathds 1}-{\mathds 1}_{[E-\delta, E+\delta]} (H)| \le \delta^{-1} |H-E|$, the norm estimate in  Theorem \ref{thm:CoulombPlasmonStates} implies that 
\begin{equation}
\| ({\mathds 1}- {\mathds 1}_{[M\epsilon_k-\delta, M\epsilon_k+\delta]} (H_{\rm eff})) \hat{\Psi}_M \| \ll 1 \quad \text{for} \quad \left|k\right|^{-1}\sqrt{k_{F}}M^{\frac{5}{2}} \ll \delta,
\end{equation}
namely  the state $\hat{\Psi}_M$ is essentially localized in the spectral space ${\mathds 1}_{[M\epsilon_k-\delta, M\epsilon_k+\delta]} (H_{\rm eff})\mathcal{H}_{N}$.  
This justifies the interpretation that $\hat{\Psi}_M$ is an ``approximate eigenfunction'' of $H_{\rm eff}$. 



\bigskip

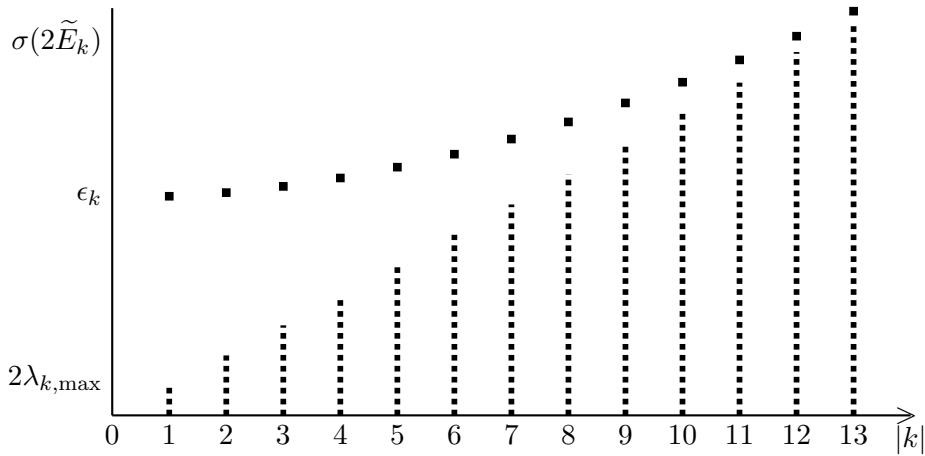
\begin{figure}[!h]
\centering
		\begin{tikzpicture}
			\draw[thick] (0,0) -- (14*0.75,0);
			
			\node [below] at (14*0.75,0) {$|k|$};
			\node at (13.92*0.75,0) {$>$};
			\node [below] at (0,0) {$0$};
			\node [below] at (1*0.75,0) {$1$};
			\node [below] at (2*0.75,0) {$2$};
			\node [below] at (3*0.75,0) {$3$};
			\node [below] at (4*0.75,0) {$4$};
			\node [below] at (5*0.75,0) {$5$};
			\node [below] at (6*0.75,0) {$6$};
			\node [below] at (7*0.75,0) {$7$};
			\node [below] at (8*0.75,0) {$8$};
			\node [below] at (9*0.75,0) {$9$};
			\node [below] at (10*0.75,0) {$10$};
			\node [below] at (11*0.75,0) {$11$};
			\node [below] at (12*0.75,0) {$12$};
			\node [below] at (13*0.75,0) {$13$};
			
			\draw[thick] (0,0) -- (0,5.4);
			\node [left] at (0,5) {$\sigma(2\widetilde{E}_k)$};
			\node [left] at (0,0.0004*7324) {$\epsilon_k$};
			\node [left] at (0,0.0004*1201) {$2\lambda_{k,{\rm max}}$};	
			
			\draw [fill] (1*0.75-0.05,0.0004*7324-0.05) rectangle ++ (0.1,0.1);   
			\draw[dotted, line width=0.75mm] (1*0.75,0) to (1*0.75,0.0004*1001);			
		
			\draw [fill] (2*0.75-0.05,0.0004*7447-0.05) rectangle ++ (0.1,0.1);   
			\draw[dotted, line width=0.75mm] (2*0.75,0) to (2*0.75,0.0004*2004);
			
			\draw [fill] (3*0.75-0.05,0.0004*7652-0.05) rectangle ++ (0.1,0.1);   
			\draw[dotted, line width=0.75mm] (3*0.75,0) to (3*0.75,0.0004*3009);
						
			\draw [fill] (4*0.75-0.05,0.0004*7935-0.05) rectangle ++ (0.1,0.1);   
			\draw[dotted, line width=0.75mm] (4*0.75,0) to (4*0.75,0.0004*4016);
	
			\draw [fill] (5*0.75-0.05,0.0004*8296-0.05) rectangle ++ (0.1,0.1);   
			\draw[dotted, line width=0.75mm] (5*0.75,0) to (5*0.75,0.0004*5025);		

			\draw [fill] (6*0.75-0.05,0.0004*8730-0.05) rectangle ++ (0.1,0.1);   
			\draw[dotted, line width=0.75mm] (6*0.75,0) to (6*0.75,0.0004*6036);

			\draw [fill] (7*0.75-0.05,0.0004*9236-0.05) rectangle ++ (0.1,0.1);   
			\draw[dotted, line width=0.75mm] (7*0.75,0) to (7*0.75,0.0004*7049);

		        \draw [fill] (8*0.75-0.05,0.0004*9808-0.05) rectangle ++ (0.1,0.1);   
			\draw[dotted, line width=0.75mm] (8*0.75,0) to (8*0.75,0.0004*8064);
			
			\draw [fill] (9*0.75-0.05,0.0004*10443-0.05) rectangle ++ (0.1,0.1);   
			\draw[dotted, line width=0.75mm] (9*0.75,0) to (9*0.75,0.0004*9081);
			
			\draw [fill] (10*0.75-0.05,0.0004*11136-0.05) rectangle ++ (0.1,0.1);  
			\draw[dotted, line width=0.75mm] (10*0.75,0) to (10*0.75,0.0004*10100);

			\draw [fill] (11*0.75-0.05,0.0004*11881-0.05) rectangle ++ (0.1,0.1);  
			\draw[dotted, line width=0.75mm] (11*0.75,0) to (11*0.75,0.0004*11121);
						
			\draw [fill] (12*0.75-0.05,0.0004*12674-0.05) rectangle ++ (0.1,0.1);  
			\draw[dotted, line width=0.75mm] (12*0.75,0) to (12*0.75,0.0004*12144);
			
			\draw [fill] (13*0.75-0.05,0.0004*13510-0.05) rectangle ++ (0.1,0.1);  
			\draw[dotted, line width=0.75mm] (13*0.75,0) to (13*0.75,0.0004*13169);

%
%
%

		\end{tikzpicture}
		
				\caption{The spectrum of $2 \widetilde{E}_{k}$ with $\hat V_k=4\pi |k|^{-2}$ and  $k_F=500$. 
				}
		\label{fig:Plasmon}
	\end{figure}

{\bf 2.}  The condition $|k| \le k_F^{\delta}$ with $\delta<1/2$ is natural since we need $|k|\ll k_F^{1/2}$ to separate the plasmon frequency from other eigenvalues of $2 \widetilde{E}_{k}$. When $|k|\sim k_F^{1/2}$, the plasmon mode merges into the continuum (the interval $[0, 2 \lambda_{k, \max}]$ containing the remaining spectrum of $2 \widetilde{E}_k$) as argued already by Bohm and Pines. See Figure \ref{fig:Plasmon} for a numerical computation of the plasmon frequency and the continuum spectrum of $2 \widetilde{E}_{k}$ when $|k|$ increases.

\bigskip

{\bf 3.} The estimate for $\epsilon_k$ is quite precise. Evidently the error term $k_F^{-1/2} |k|^4$ is much smaller than  $k_F^{3/2}$ when $|k|\ll k_F^{1/2}$; moreover  
\begin{equation}
\epsilon_{k}=\sqrt{8\left\langle v_{k},h_{k}v_{k}\right\rangle +4\frac{\left\langle v_{k},h_{k}^{3}v_{k}\right\rangle }{\left\langle v_{k},h_{k}v_{k}\right\rangle }}+o\left(1\right)
\end{equation}
for $\left|k\right|\ll k_{F}^{\frac{1}{8}}$. To make connections to the physics literature, we note that replacing the underlying
Riemann sums by integrals (and keeping only the leading part of $\left\langle v_{k},h_{k}^{3}v_{k}\right\rangle $), and setting $g = 4 \pi e^2$ and $n=\frac{N}{\mathcal{V}} = \frac{4\pi}{3}k_F^3$ we find (with $\frac{\hbar^2}{2 m} = 1$)
\begin{equation} \label{eq:eps-k-af-thm}
\epsilon_{k}\approx\sqrt{\frac{4 \pi e^2}{3\pi^2}k_{F}^{3}+\frac{12}{5}k_{F}^{2}\left|k\right|^{2}} \approx \hbar \omega_0 + \frac{\frac{6}{5} k_F^2 \vert k \vert^2}{\hbar \omega_0}
= \hbar \left( \omega_0 + \frac{3}{10} \frac{v_F^2}{\omega_0} \vert k \vert^2 \right),
\end{equation}
where $\omega_{0}= \sqrt{\frac{4\pi ne^{2}}{m}}$ is the plasmon frequency and $v_{F}= 2 \hbar^{-1} k_F$ is the Fermi velocity. This describes a plasmon dispersion relation of
\begin{equation} 
\omega(k) \approx \omega_0 + \frac{3}{10} \frac{v_F^2}{\omega_0} \vert k \vert^2,
\end{equation}
which is in agreement with \cite[Eq. (3.90c)]{Pines-99}, \cite[Eq. (5.19) ]{Pines-53} and \cite[Eq. (15.60)]{Fetter-Walecka-71}. See Section \ref{sec:Conclusion} for a detailed explanation of \eqref{eq:eps-k-af-thm}.


\bigskip

{\bf 4.} In the mean-field regime, where $V=k_F^{-1}W$ with a fixed potential $W$, the bosonic collective excitations were discussed in  \cite{BNPSS-22} on the dynamics and in \cite{Benedikter-20} on the spectrum (see e.g. \cite[Eq.  (3.38)]{Benedikter-20} for an analogue of \eqref{eq:eps-k-af-thm}). In this case, the  separation of  the plasmon frequency holds in a weak sense: although the largest eigenvalue of $2\widetilde{E}_{k}$ are within the same order of magnitude of many other eigenvalues, i.e. of order $k_F$,  the distance from from the plasmon frequency to the next-highest one is also of order $k_F$ while the gaps between other eigenvalues are at most $O(|k|)$ (recall that we are interested in the case $|k|\ll k_F^{1/2}$). This assertion follows easily from the same argument leading to \eqref{eq:eps-lower-bound-first}. 

\bigskip

In contrast, in the present work we focus on the more physical regime where $V$ is independent of $k_F$. As we go beyond the mean-field regime, the largest eigenvalue is much larger than the others, and the genuninely large gap of the spectrum ensures the almost-delocalization of the eigenfunction, which is important for our estimate. 

\bigskip

{\bf 5.} Our analysis can be extended to all potentials satisfying $\hat V_k \ge 0$ and $\sum_{k\in \mathbb{Z}^{3}_*} \hat V_k^2 <\infty$. To be precise, 
for any $k\in\mathbb{Z}_{\ast}^{3}$ and $M\in\mathbb{N}$ such that $\hat{V}_{k}\gg k_{F}^{-1}$ and $1\le M\ll k_{F}\left|k\right|^{\frac{1}{2}}$, if we take $\epsilon_k, \varphi_k$ and $\hat \Psi_M$ as in Theorem \ref{thm:CoulombPlasmonStates}, then we have the norm estimate 
\begin{equation} \label{eq:intro-gen-1}
\left\Vert \left(H_{\mathrm{eff}}-M\epsilon_{k}\right)\hat\Psi_M\right\Vert \leq \frac{C}{\sqrt{k_{F}}\left|k\right|}\sqrt{\sum_{l\in2B_{F}}\min\left\{ 1,k_{F}\hat{V}_{l}\right\} \hat{V}_{l}\left|l\right|^{2}+Ck_{F}^{3}\sum_{l\in\mathbb{Z}^{3}\backslash2B_{F}}\hat{V}_{l}^{2}}M^{\frac{5}{2}}.
\end{equation}
Note that in the case of the Coulomb potential $\hat{V}_{k}=g\left|k\right|^{-2}$
we may explicitly estimate
\begin{align}
\sum_{l\in2B_{F}}\min\left\{ 1,k_{F}\hat{V}_{l}\right\} \hat{V}_{l}\left|l\right|^{2} & \sim\int_{0}^{\sqrt{k_{F}}}r^{2}dr+k_{F}\int_{\sqrt{k_{F}}}^{2k_{F}}\frac{1}{r^{2}}r^{2}dr\leq k_{F}^{\frac{3}{2}}+k_{F}k_{F}\leq k_{F}^{2},\nonumber \\
k_{F}^{3}\sum_{l\in\mathbb{Z}^{3}\backslash2B_{F}}\hat{V}_{l}^{2} & \sim k_{F}^{3}\int_{2k_{F}}^{\infty}\frac{1}{r^{4}}r^{2}dr\leq k_{F}^{3}k_{F}^{-1}=k_{F}^{2},
\end{align}
and hence \eqref{eq:intro-gen-1}  boils down to the norm estimate in Theorem \ref{thm:CoulombPlasmonStates}. We refer to Section \ref{sec:Conclusion} for further explanation of \eqref{eq:intro-gen-1}. 

\bigskip

{\bf Outline of the proof:} The main mathematical difficulty of the proof lies on the fact that $H_{\rm eff}$ is not a bosonic operator. More precisely, the operators $b_{k}\left(\varphi\right)$ only satisfies the CCR in a weak sense, and controlling the exchange terms (the error terms from the CCR) requires a careful analysis.  In particular, estimating the norm of the approximate eigenstate $b_{k}^{\ast}\left(\varphi\right)^{M}\psi_{\FS}$ is already nontrivial, and this will be done in Section \ref{sec:AnalysisoftheTrialState}, together with an analysis of the action of $H_{\rm eff}$ on this state. Until this point, we keep the analysis general and do not use any properties of the one-body operators $\widetilde{E}_k$ and $h_k$ in the definition of $H_{\rm eff}$.   These one-body operators will be analyzed in detail in Section \ref{sec:EstimationofOne-BodyQuantities}. Finally, we conclude the proof of the main theorem in Section \ref{sec:Conclusion}. 

\bigskip

{\bf Acknowledgements:} We thank the referees for helpful suggestions. MRC and PTN acknowledge the support from the
Deutsche Forschungsgemeinschaft (DFG project Nr. 426365943).

\section{Analysis of the Approximate Eigenstates} \label{sec:AnalysisoftheTrialState}

Let $k\in\mathbb{Z}_{\ast}^{3}$ be given and let $\varphi\in\ell^{2}\left(L_{k}\right)$
be the normalized eigenstate of $2\widetilde{E}_{k}$ corresponding
to the greatest eigenvalue $\epsilon_{k}$. For $M\in\mathbb{N}_{0}$
we define a state $\Psi_{M}\in\left\{ \mathcal{N}_{E}=M\right\} $
by
\begin{equation}
\Psi_{M}=b_{k}^{\ast}\left(\varphi\right)^{M}\psi_{\FS}. 
\end{equation}

In this section, we estimate the norms of  $\Psi_M$ and $(H_{\rm eff}-M \epsilon_k) \Psi_M/\|\Psi_M\|$; the main results are stated in Proposition \ref{prop:NormLowerBound} and Proposition \ref{prop:SpectralEstimate}, respectively. 

Before going to the two corresponding subsections, let us recall some basic commutator computations. 
First, we recall that the generalized excitation operators, given by  
\begin{equation}
b_{k}\left(\varphi\right)=\sum_{p\in L_{k}}\left\langle \varphi,e_{p}\right\rangle b_{k,p},\quad b_{k}^{\ast}\left(\varphi\right)=\sum_{p\in L_{k}}\left\langle e_{p},\varphi\right\rangle b_{k,p}^{\ast},
\end{equation}
with $b_{k,p}=c_{p-k}^{\ast}c_{p}$,  obey the commutation relations
\begin{align}
\left[b_{k}\left(\varphi\right),b_{l}\left(\psi\right)\right] & =\left[b_{k}^{\ast}\left(\varphi\right),b_{l}^{\ast}\left(\psi\right)\right]=0\label{eq:Quasi-BosonicCCR}\\
\left[b_{k}\left(\varphi\right),b_{l}^{\ast}\left(\psi\right)\right] & =\delta_{k,l}\left\langle \varphi,\psi\right\rangle +\varepsilon_{k,l}\left(\varphi;\psi\right)\nonumber 
\end{align}
with
\begin{equation}
\varepsilon_{k,l}\left(\varphi;\psi\right)=-\sum_{p\in L_{k}}\sum_{q\in L_{l}}\left\langle \varphi,e_{p}\right\rangle \left\langle e_{q},\psi\right\rangle \left(\delta_{p,q}c_{q-l}c_{p-k}^{\ast}+\delta_{p-k,q-l}c_{q}^{\ast}c_{p}\right).
\end{equation}
For use below we calculate the commutator $\left[\varepsilon_{l,k}\left(\phi;\varphi\right),b_{k}^{\ast}\left(\psi\right)\right]$:
As
\begin{align}
\left[\delta_{p,q}c_{q-k}c_{p-l}^{\ast}+\delta_{p-l,q-k}c_{q}^{\ast}c_{p},b_{k,r}^{\ast}\right] & =\delta_{p,q}\left[c_{q-k}c_{p-l}^{\ast},c_{r}^{\ast}c_{r-k}\right]+\delta_{p-l,q-k}\left[c_{q}^{\ast}c_{p},c_{r}^{\ast}c_{r-k}\right]\nonumber \\
 & =\delta_{p,q}c_{r}^{\ast}c_{q-k}\left\{ c_{p-l}^{\ast},c_{r-k}\right\} +\delta_{p-l,q-k}c_{q}^{\ast}\left\{ c_{p},c_{r}^{\ast}\right\} c_{r-k}\nonumber \\
 & =\delta_{p,q}\delta_{p-l,r-k}c_{r}^{\ast}c_{q-k}+\delta_{p,r}\delta_{p-l,q-k}c_{q}^{\ast}c_{r-k}
\end{align}
for $p\in L_{l}$ and $q,r\in L_{k}$, we find
\begin{align}
\left[\varepsilon_{l,k}\left(\phi;\varphi\right),b_{k}^{\ast}\left(\psi\right)\right] & =-\sum_{p\in L_{l}}\sum_{q,r\in L_{k}}\left\langle \phi,e_{p}\right\rangle \left\langle e_{q},\varphi\right\rangle \left\langle e_{r},\psi\right\rangle \left[\delta_{p,q}c_{q-k}c_{p-l}^{\ast}+\delta_{p-l,q-k}c_{q}^{\ast}c_{p},b_{k,r}^{\ast}\right]\nonumber \\
 & =-\sum_{p\in L_{k}\cap L_{l}}\sum_{r\in L_{k}}\left\langle \phi,e_{p}\right\rangle \left\langle e_{p},\varphi\right\rangle \left\langle e_{r},\psi\right\rangle \delta_{p-l,r-k}c_{r}^{\ast}c_{p-k}\nonumber \\
 &\quad -\sum_{p\in L_{k}\cap L_{l}}\sum_{q\in L_{k}}\left\langle \phi,e_{p}\right\rangle \left\langle e_{q},\varphi\right\rangle \left\langle e_{p},\psi\right\rangle \delta_{p-l,q-k}c_{q}^{\ast}c_{p-k}\label{eq:ExchangeCorrectionCommutator}\\
 & =-\sum_{p\in L_{k}\cap L_{l}}\sum_{q\in L_{k}}\left\langle \phi,e_{p}\right\rangle \left(\left\langle e_{p},\varphi\right\rangle \left\langle e_{q},\psi\right\rangle +\left\langle e_{q},\varphi\right\rangle \left\langle e_{p},\psi\right\rangle \right)\delta_{p-l,q-k}c_{q}^{\ast}c_{p-k}\nonumber \\
 & =-\sum_{p\in L_{k}\cap L_{l}}\sum_{q\in L_{k}}\delta_{p-l,q-k}\left\langle \phi,e_{p}\right\rangle \left(\left\langle e_{p},\varphi\right\rangle \left\langle e_{q},\psi\right\rangle +\left\langle e_{q},\varphi\right\rangle \left\langle e_{p},\psi\right\rangle \right)b_{2k-l,q}^{\ast}.\nonumber 
\end{align}
In particular
\begin{align}
\left[\varepsilon_{k,k}\left(\phi;\varphi\right),b_{k}^{\ast}\left(\psi\right)\right] & =-\sum_{p\in L_{k}}\sum_{q\in L_{k}}\delta_{p-k,q-k}\left\langle \phi,e_{p}\right\rangle \left(\left\langle e_{p},\varphi\right\rangle \left\langle e_{q},\psi\right\rangle +\left\langle e_{q},\varphi\right\rangle \left\langle e_{p},\psi\right\rangle \right)b_{k,q}^{\ast}\label{eq:ExchangeCorrectionCommutatorkeql} \nonumber\\
 & =-2\sum_{p\in L_{k}}\left\langle \phi,e_{p}\right\rangle \left\langle e_{p},\varphi\right\rangle \left\langle e_{p},\psi\right\rangle b_{k,p}^{\ast}. 
\end{align}

\subsection{Estimating the Norm of $\Psi_{M}$}

In this subsection we will prove the following: 

\begin{prop}
\label{prop:NormLowerBound}It holds that
\[
M! \ge \left\Vert \Psi_{M}\right\Vert ^{2}\geq M!\left(1-\frac{M\left(M-1\right)}{2}\left\Vert \varphi\right\Vert _{6}^{3}\right).
\]
\end{prop}

Below we will see that $\varphi$ is ``almost completely delocalized'',
i.e. $\left|\left\langle e_{p},\varphi\right\rangle \right|\sim\dim\left(\ell^{2}\left(L_{k}\right)\right)^{-\frac{1}{2}}=\left|L_{k}\right|^{-\frac{1}{2}}$,
whence
\begin{equation}
\left\Vert \varphi\right\Vert _{6}^{3}\sim\sqrt{\sum_{p\in L_{k}}\frac{1}{\left|L_{k}\right|^{3}}}=\left|L_{k}\right|^{-1}
\end{equation}
and so the proposition implies that
\begin{equation}
\left\Vert \Psi_{M}\right\Vert ^{2}\geq C\left(M!\right),\quad M\ll\left|L_{k}\right|^{\frac{1}{2}}\sim k_{F}\left|k\right|^{\frac{1}{2}}.
\end{equation}

We note the following general estimates:
\begin{lem}
\label{prop:ExcitationSumEstimate}Let $\left(\phi_{k}\right)_{k\in\mathbb{Z}_{\ast}^{3}}$
be a collection of vectors $\phi_{k}\in\ell^{2}\left(L_{k}\right)$.
Then for any $\Psi\in\mathcal{H}_{N}$
\begin{align*}
\left\Vert \sum_{k\in\mathbb{Z}_{\ast}^{3}}b_{k}\left(\phi_{k}\right)\Psi\right\Vert  & \leq\sqrt{\sum_{k\in\mathbb{Z}_{\ast}^{3}}\left\Vert \phi_{k}\right\Vert ^{2}}\left\Vert \mathcal{N}_{E}^{\frac{1}{2}}\Psi\right\Vert, \\
\left\Vert \sum_{k\in\mathbb{Z}_{\ast}^{3}}b_{k}^{\ast}\left(\phi_{k}\right)\Psi\right\Vert  & \leq\sqrt{\sum_{k\in\mathbb{Z}_{\ast}^{3}}\left\Vert \phi_{k}\right\Vert ^{2}}\left\Vert \left(\mathcal{N}_{E}+1\right)^{\frac{1}{2}}\Psi\right\Vert .
\end{align*}
\end{lem}
\textbf{Proof:} For $\sum_{k\in\mathbb{Z}_{\ast}^{3}}b_{k}\left(\phi_{k}\right)\Psi$
we estimate
\begin{align}
 & \left\Vert \sum_{k\in\mathbb{Z}_{\ast}^{3}}b_{k}\left(\phi_{k}\right)\Psi\right\Vert =\left\Vert \sum_{k\in\mathbb{Z}_{\ast}^{3}}\sum_{p\in L_{k}}\left\langle \phi_{k},e_{p}\right\rangle c_{p-k}^{\ast}c_{p}\Psi\right\Vert \\
 = & \left\Vert \sum_{p\in B_{F}^{c}}\sum_{k\in\mathbb{Z}_{\ast}^{3}}1_{L_{k}}\left(p\right)\left\langle \phi_{k},e_{p}\right\rangle c_{p-k}^{\ast}c_{p}\Psi\right\Vert \leq\sum_{p\in B_{F}^{c}}\left\Vert \left(\sum_{k\in\mathbb{Z}_{\ast}^{3}}1_{L_{k}}\left(p\right)\left\langle \phi_{k},e_{p}\right\rangle c_{p-k}^{\ast}\right)c_{p}\Psi\right\Vert \nonumber \\
  \leq & \sum_{p\in B_{F}^{c}}\sqrt{\sum_{k\in\mathbb{Z}_{\ast}^{3}}1_{L_{k}}\left(p\right)\left|\left\langle \phi_{k},e_{p}\right\rangle \right|^{2}}\left\Vert c_{p}\Psi\right\Vert \leq\sqrt{\sum_{p\in B_{F}^{c}}\sum_{k\in\mathbb{Z}_{\ast}^{3}}1_{L_{k}}\left(p\right)\left|\left\langle \phi_{k},e_{p}\right\rangle \right|^{2}}\sqrt{\sum_{p\in B_{F}^{c}}\left\Vert c_{p}\Psi\right\Vert ^{2}}\nonumber \\
  =&\sqrt{\sum_{k\in\mathbb{Z}_{\ast}^{3}}\left\Vert \phi_{k}\right\Vert ^{2}}\sqrt{\left\langle \Psi,\mathcal{N}_{E}\Psi\right\rangle }=\sqrt{\sum_{k\in\mathbb{Z}_{\ast}^{3}}\left\Vert \phi_{k}\right\Vert ^{2}}\left\Vert \mathcal{N}_{E}^{\frac{1}{2}}\Psi\right\Vert \nonumber 
\end{align}
by the usual fermionic estimate $\left\Vert \sum_{p}a_{p}c_{p}^{\ast}\right\Vert _{\mathrm{Op}}\leq\sqrt{\sum_{p}\left|a_{p}\right|^{2}}$.
For the second estimate we note that by the quasi-bosonic commutation
relations of equation (\ref{eq:Quasi-BosonicCCR})
\begin{align}
\left[\sum_{k\in\mathbb{Z}_{\ast}^{3}}b_{k}\left(\phi_{k}\right),\sum_{l\in\mathbb{Z}_{\ast}^{3}}b_{l}^{\ast}\left(\phi_{l}\right)\right] & =\sum_{k,l\in\mathbb{Z}_{\ast}^{3}}\delta_{k,l}\left\langle \phi_{k},\phi_{l}\right\rangle +\sum_{k,l\in\mathbb{Z}_{\ast}^{3}}\varepsilon_{k,l}\left(\phi_{k};\phi_{l}\right)\\
 & =\sum_{k\in\mathbb{Z}_{\ast}^{3}}\left\Vert \phi_{k}\right\Vert ^{2}+\sum_{k,l\in\mathbb{Z}_{\ast}^{3}}\varepsilon_{k,l}\left(\phi_{k};\phi_{l}\right)\nonumber 
\end{align}
whence the second estimate will follow from the first provided $\sum_{k,l\in\mathbb{Z}_{\ast}^{3}}\varepsilon_{k,l}\left(\phi_{k};\phi_{l}\right)\leq0$.
This is indeed the case since by definition
\begin{equation}
\sum_{k,l\in\mathbb{Z}_{\ast}^{3}}\varepsilon_{k,l}\left(\phi_{k};\phi_{l}\right)=-\sum_{k,l\in\mathbb{Z}_{\ast}^{3}}\sum_{p\in L_{k}}\sum_{q\in L_{l}}\left\langle \phi_{k},e_{p}\right\rangle \left\langle e_{q},\phi_{l}\right\rangle \left(\delta_{p,q}c_{q-l}c_{p-k}^{\ast}+\delta_{p-k,q-l}c_{q}^{\ast}c_{p}\right)
\end{equation}
which factorizes as the negative of a sum of squares: Firstly
\begin{align}
 & \quad\,\sum_{k,l\in\mathbb{Z}_{\ast}^{3}}\sum_{p\in L_{k}}\sum_{q\in L_{l}}\left\langle \phi_{k},e_{p}\right\rangle \left\langle e_{q},\phi_{l}\right\rangle \delta_{p,q}c_{q-l}c_{p-k}^{\ast}=\sum_{k,l\in\mathbb{Z}_{\ast}^{3}}\sum_{p\in L_{k}\cap L_{l}}\left\langle \phi_{k},e_{p}\right\rangle \left\langle e_{p},\phi_{l}\right\rangle c_{p-l}c_{p-k}^{\ast}\nonumber \\
 & =\sum_{p\in B_{F}^{c}}\left(\sum_{l\in\mathbb{Z}_{\ast}^{3}}1_{L_{l}}\left(p\right)\left\langle e_{p},\phi_{l}\right\rangle c_{p-l}\right)\left(\sum_{k\in\mathbb{Z}_{\ast}^{3}}1_{k}\left(p\right)\left\langle e_{p},\phi_{k}\right\rangle c_{p-k}\right)^{\ast}.
\end{align}
Similarly
\begin{align}
 & \quad\;\sum_{k,l\in\mathbb{Z}_{\ast}^{3}}\sum_{p\in L_{k}}\sum_{q\in L_{l}}\left\langle \phi_{k},e_{p}\right\rangle \left\langle e_{q},\phi_{l}\right\rangle \delta_{p-k,q-l}c_{q}^{\ast}c_{p}\nonumber \\
 & =\sum_{k,l\in\mathbb{Z}_{\ast}^{3}}\sum_{p\in\left(L_{k}-k\right)}\sum_{q\in\left(L_{l}-l\right)}\left\langle \phi_{k},e_{p+k}\right\rangle \left\langle e_{q+l},\phi_{l}\right\rangle \delta_{p,q}c_{q+l}^{\ast}c_{p+k}\\
 & =\sum_{k,l\in\mathbb{Z}_{\ast}^{3}}\sum_{p\in\left(L_{k}-k\right)\cap\left(L_{l}-l\right)}\left\langle \phi_{k},e_{p+k}\right\rangle \left\langle e_{p+l},\phi_{l}\right\rangle c_{p+l}^{\ast}c_{p+k}\nonumber \\
 & =\sum_{p\in B_{F}}\left(\sum_{l\in\mathbb{Z}_{\ast}^{3}}1_{L_{l}-l}\left(p\right)\left\langle \phi_{l},e_{p+l}\right\rangle c_{p+l}^{\ast}\right)^{\ast}\left(\sum_{k\in\mathbb{Z}_{\ast}^{3}}1_{L_{k}-k}\left(p\right)\left\langle \phi_{k},e_{p+k}\right\rangle c_{p+k}\right).\nonumber 
\end{align}

$\hfill\square$

\textbf{Proof of Proposition \ref{prop:NormLowerBound} (Upper bound):} For any $\Psi\in\mathcal{H}_{N}$ and $\phi\in\ell^{2}\left(L_{k}\right)$
it holds that
\begin{equation}
\left\Vert b_{k}^{\ast}\left(\phi\right)\Psi\right\Vert \leq\left\Vert \phi\right\Vert \sqrt{\left\langle \Psi,\left(\mathcal{N}_{E}+1\right)\Psi\right\rangle}
\end{equation}
(this is a special case of Lemma \ref{prop:ExcitationSumEstimate}). In particular, since $\varphi\in \ell^{2}\left(L_{k}\right)$ is normalized and $\Psi_{M}\in\left\{ \mathcal{N}_{E}=M\right\}$ we have
\begin{equation}
\left\Vert \Psi_{M}\right\Vert ^{2}=\left\Vert b^{\ast}\left(\varphi\right)\Psi_{M-1}\right\Vert ^{2}\leq\left\Vert \varphi\right\Vert ^{2}\left\langle \Psi_{M-1},\left(\mathcal{N}_{E}+1\right)\Psi_{M-1}\right\rangle =M\left\Vert \Psi_{M-1}\right\Vert ^{2}
\end{equation}
whence $\left\Vert \Psi_{M}\right\Vert ^{2}\leq M!\left\Vert \psi_{\FS}\right\Vert ^{2}=M!$. 

$\hfill\square$

Obtaining the lower bound will require some additional work. 
We note the following:
\begin{lem}
\label{lemma:CommutationLemma}For any $l\in\mathbb{Z}_{\ast}^{3}$
and $\phi\in\ell^{2}\left(L_{l}\right)$ it holds that
\[
b_{l}\left(\phi\right)\Psi_{M}=\delta_{k,l}M\left\langle \phi,\varphi\right\rangle \Psi_{M-1}+\frac{M\left(M-1\right)}{2}\left[\varepsilon_{l,k}\left(\phi;\varphi\right),b_{k}^{\ast}\left(\varphi\right)\right]\Psi_{M-2}.
\]
In particular, for $k=l$,
\[
b_{k}\left(\phi\right)\Psi_{M}=M\left\langle \phi,\varphi\right\rangle \Psi_{M-1}-M\left(M-1\right)\left(\sum_{p\in L_{k}}\left\langle \phi,e_{p}\right\rangle \left\langle e_{p},\varphi\right\rangle ^{2}b_{k,p}^{\ast}\right)\Psi_{M-2}.
\]
\end{lem}
\textbf{Proof:} We calculate
\begin{align} \label{eq:4.12}
\left[b_{l}\left(\phi\right),b_{k}^{\ast}\left(\varphi\right)^{M}\right] & =\sum_{j=1}^{M}b_{k}^{\ast}\left(\varphi\right)^{M-j}\left[b_{l}\left(\phi\right),b_{k}^{\ast}\left(\varphi\right)\right]b_{k}^{\ast}\left(\varphi\right)^{j-1}\nonumber \\
 & =\delta_{k,l}\left\langle \phi,\varphi\right\rangle \sum_{j=1}^{M}b_{k}^{\ast}\left(\varphi\right)^{M-1}+\sum_{j=1}^{M}b_{k}^{\ast}\left(\varphi\right)^{M-j}\varepsilon_{l,k}\left(\phi;\varphi\right)b_{k}^{\ast}\left(\varphi\right)^{j-1}\\
 & =\delta_{k,l}M\left\langle \phi,\varphi\right\rangle b_{k}^{\ast}\left(\varphi\right)^{M-1}+Mb_{k}^{\ast}\left(\varphi\right)^{M-1}\varepsilon_{l,k}\left(\phi;\varphi\right)\nonumber \\
 & +\sum_{j=1}^{M}\sum_{j'=1}^{j-1}b_{k}^{\ast}\left(\varphi\right)^{M-j}b_{k}^{\ast}\left(\varphi\right)^{j-1-j'}\left[\varepsilon_{l,k}\left(\phi;\varphi\right),b_{k}^{\ast}\left(\varphi\right)\right]b_{k}^{\ast}\left(\varphi\right)^{j'-1}.\nonumber 
\end{align}
Here the third equation in \eqref{eq:4.12} is obtained by iterating the second one and commuting the operator $b_k^*(\varphi)$ to the left. Note that it follows from equation (\ref{eq:ExchangeCorrectionCommutator})
that $\left[\varepsilon_{l,k}\left(\phi;\varphi\right),b_{k}^{\ast}\left(\varphi\right)\right]$
commutes with $b_{k}^{\ast}\left(\varphi\right)$. As $b_{l}\left(\phi\right)\psi_{\FS}=0=\varepsilon_{l,k}\left(\phi;\varphi\right)\psi_{\FS}$
we thus find by applying \eqref{eq:4.12} to $\psi_{\rm FS}$ that
\begin{align}
b_{l}\left(\phi\right)\Psi_{M} & =\delta_{k,l}M\left\langle \phi,\varphi\right\rangle b_{k}^{\ast}\left(\varphi\right)^{M-1}\psi_{\FS}+\left(\sum_{j=1}^{M}\sum_{j'=1}^{j-1}1\right)\left[\varepsilon_{l,k}\left(\phi;\varphi\right),b_{k}^{\ast}\left(\varphi\right)\right]b_{k}^{\ast}\left(\varphi\right)^{M-2}\psi_{\rm FS}\nonumber \\
 & =\delta_{k,l}M\left\langle \phi,\varphi\right\rangle \Psi_{M-1}+\frac{M\left(M-1\right)}{2}\left[\varepsilon_{l,k}\left(\phi;\varphi\right),b_{k}^{\ast}\left(\varphi\right)\right]\Psi_{M-2}.
\end{align}
The $k=l$ case follows by inserting equation (\ref{eq:ExchangeCorrectionCommutatorkeql}).

$\hfill\square$

\textbf{Proof of Proposition \ref{prop:NormLowerBound} (Lower bound):} We define $\varphi^{\left(3\right)}\in\ell^{2}\left(L_{k}\right)$
by
\begin{equation}
\varphi^{\left(3\right)}=\sum_{p\in L_{k}}\left|\left\langle e_{p},\varphi\right\rangle \right|^{2}\left\langle e_{p},\varphi\right\rangle e_{p}.
\end{equation}
We then see by Lemma \ref{lemma:CommutationLemma} that $\left\Vert \Psi_{M}\right\Vert ^{2}$ obeys
\begin{align}
\left\Vert \Psi_{M}\right\Vert ^{2} & =\left\langle \Psi_{M-1},b_{k}\left(\varphi\right)\Psi_{M}\right\rangle \label{eq:PsiMNormRecursion}\\
 & =\left\langle \Psi_{M-1},M\left\langle \varphi,\varphi\right\rangle \Psi_{M-1}-M\left(M-1\right)\left(\sum_{p\in L_{k}}\left\langle \varphi,e_{p}\right\rangle \left\langle e_{p},\varphi\right\rangle ^{2}b_{k,p}^{\ast}\right)\Psi_{M-2}\right\rangle \nonumber \\
 & =M\left\Vert \Psi_{M-1}\right\Vert ^{2}-M\left(M-1\right)\left\langle \Psi_{M-1},b_{k}^{\ast}\left(\varphi^{\left(3\right)}\right)\Psi_{M-2}\right\rangle .\nonumber 
\end{align}
From this we can deduce the desired lower bound by induction. For $M=0,1$ we have equality.
Suppose that case $M-1$ holds. Then
\begin{align}
\left\Vert \Psi_{M}\right\Vert ^{2} & =M\left\Vert \Psi_{M-1}\right\Vert ^{2}-M\left(M-1\right)\left\langle \Psi_{M-1},b_{k}^{\ast}\left(\varphi^{\left(3\right)}\right)\Psi_{M-2}\right\rangle \nonumber \\
 & \geq M\left\Vert \Psi_{M-1}\right\Vert ^{2}-M\left(M-1\right)\left\Vert \Psi_{M-1}\right\Vert \left\Vert b_{k}^{\ast}\left(\varphi^{\left(3\right)}\right)\Psi_{M-2}\right\Vert \nonumber \\
 & \geq M\left\Vert \Psi_{M-1}\right\Vert ^{2}-M\left(M-1\right)^{\frac{3}{2}}\left\Vert \varphi^{\left(3\right)}\right\Vert \left\Vert \Psi_{M-1}\right\Vert \left\Vert \Psi_{M-2}\right\Vert \\
 & \geq M\left(\left(M-1\right)!\left(1-\frac{\left(M-1\right)\left(M-2\right)}{2}\left\Vert \varphi\right\Vert _{6}^{3}\right)\right)-M\left(M-1\right)\left\Vert \varphi\right\Vert _{6}^{3}\left(M-1\right)!\nonumber \\
 & =M!\left(\left(1-\frac{\left(M-1\right)\left(M-2\right)}{2}\left\Vert \varphi\right\Vert _{6}^{3}\right)-\left(M-1\right)\left\Vert \varphi\right\Vert _{6}^{3}\right)\nonumber \\
 & =M!\left(1-\frac{M\left(M-1\right)}{2}\left\Vert \varphi\right\Vert _{6}^{3}\right)\nonumber 
\end{align}
where we recognized that
\begin{equation}
\left\Vert \varphi^{\left(3\right)}\right\Vert =\sqrt{\sum_{p\in L_{k}}\left|\left\langle e_{p},\varphi\right\rangle \right|^{6}}=\left\Vert \varphi\right\Vert _{6}^{3}.
\end{equation}
The proof of Proposition \ref{prop:NormLowerBound} is complete. 

$\hfill\square$

\subsection{Action of the Effective Hamiltonian on $\Psi_{M}$}

We now consider the action of
\begin{align}
H_{\mathrm{eff}} & =H_{\mathrm{kin}}^{\prime}+2\sum_{k\in\mathbb{Z}_{\ast}^{3}}\sum_{p,q\in L_{k}}\left\langle e_{p},\left(\widetilde{E}_{k}-h_{k}\right)e_{q}\right\rangle b_{k,p}^{\ast}b_{k,q}\\
 & =H_{\mathrm{kin}}^{\prime}+2\sum_{k\in\mathbb{Z}_{\ast}^{3}}\sum_{p\in L_{k}}b_{k}^{\ast}\left(\left(\widetilde{E}_{k}-h_{k}\right)e_{p}\right)b_{k,p}=:H_{\mathrm{kin}}^{\prime}+H_{\mathrm{QB}}\nonumber 
\end{align}
on $\Psi_{M}$, and in doing so prove the following:
\begin{prop}
\label{prop:SpectralEstimate}For $\hat{\Psi}_{M}=\left\Vert \Psi_{M}\right\Vert ^{-1}\Psi_{M}$
it holds that
\[
\left\Vert \left(H_{\mathrm{eff}}-M\epsilon_{k}\right)\hat{\Psi}_{M}\right\Vert \leq\frac{2\left\Vert \varphi\right\Vert _{\infty}^{2}\sqrt{\sum_{l\in\mathbb{Z}_{\ast}^{3}}\left\Vert \widetilde{E}_{l}-h_{l}\right\Vert _{\mathrm{HS}}^{2}}}{\sqrt{1-M^{2}\left\Vert \varphi\right\Vert _{6}^{3}}}M^{\frac{5}{2}}.
\]
\end{prop}

We start with 

\begin{lem}
\label{prop:4.23} We have
\begin{equation}
\left\Vert \left(H_{\mathrm{eff}}-M\epsilon_{k}\right)\Psi_{M}\right\Vert \leq\frac{M\left(M-1\right)}{2}\left\Vert \mathcal{E} \Psi_{M-2}\right\Vert \label{eq:SpectrumEstimate}
\end{equation}
where
\begin{equation}
\mathcal{E}=\sum_{p,q\in L_{k}}\left\langle e_{p},\varphi\right\rangle \left\langle e_{q},\varphi\right\rangle \left(\sum_{l\in\mathbb{Z}_{\ast}^{3}}\delta_{p-l,q-k}1_{L_{l}}\left(p\right)b_{l}^{\ast}\left(A_{l}e_{p}\right)\right)c_{q}^{\ast}c_{p-k}.\label{eq:calEAlternateForm}
\end{equation}
with $A_{l}=2\left(\widetilde{E}_{l}-h_{l}\right)$. 
\end{lem}
\textbf{Proof:}
From the first identity of equation \eqref{eq:intro-Hkin-com} it follows that
\begin{equation}
\left[H_{\mathrm{kin}}^{\prime},b_{k}^{\ast}\left(\varphi\right)\right]=b_{k}^{\ast}\left(2h_{k}\varphi\right)
\end{equation}
whence
\begin{equation}
\left[H_{\mathrm{kin}}^{\prime},b_{k}^{\ast}\left(\varphi\right)^{M}\right]=Mb_{k}^{\ast}\left(2h_{k}\varphi\right)b_{k}^{\ast}\left(\varphi\right)^{M-1},
\end{equation}
implying that
\begin{equation}
H_{\mathrm{kin}}^{\prime}\Psi_{M}=Mb_{k}^{\ast}\left(2h_{k}\varphi\right)\Psi_{M-1}.
\end{equation}
For $H_{\mathrm{QB}}$ we have by Lemma \ref{lemma:CommutationLemma}
that (abbreviating $A_{l}=2\left(\widetilde{E}_{l}-h_{l}\right)$)
\begin{align}
& H_{\mathrm{QB}}\Psi_{M} \nonumber\\
& =\sum_{l\in\mathbb{Z}_{\ast}^{3}}\sum_{p\in L_{l}}b_{l}^{\ast}\left(A_{l}e_{p}\right)\left(\delta_{k,l}M\left\langle e_{p},\varphi\right\rangle \Psi_{M-1}+\frac{M\left(M-1\right)}{2}\left[\varepsilon_{l,k}\left(e_{p};\varphi\right),b_{k}^{\ast}\left(\varphi\right)\right]\Psi_{M-2}\right) \nonumber\\
 & =M\sum_{p\in L_{k}}b_{k}^{\ast}\left(A_{k}e_{p}\right)\left\langle e_{p},\varphi\right\rangle \Psi_{M-1} \nonumber\\
 &\quad +\frac{M\left(M-1\right)}{2}\sum_{l\in\mathbb{Z}_{\ast}^{3}}\sum_{p\in L_{l}}b_{l}^{\ast}\left(A_{l}e_{p}\right)\left[\varepsilon_{l,k}\left(e_{p};\varphi\right),b_{k}^{\ast}\left(\varphi\right)\right]\Psi_{M-2}\nonumber \\
 & =Mb_{k}^{\ast}\left(A_{k}\varphi\right)\Psi_{M-1}+\frac{M\left(M-1\right)}{2}\sum_{l\in\mathbb{Z}_{\ast}^{3}}\sum_{p\in L_{l}}b_{l}^{\ast}\left(A_{l}e_{p}\right)\left[\varepsilon_{l,k}\left(e_{p};\varphi\right),b_{k}^{\ast}\left(\varphi\right)\right]\Psi_{M-2}.\nonumber 
\end{align}
In all then
\begin{equation}
H_{\mathrm{eff}}\Psi_{M}=Mb_{k}^{\ast}\left(2\widetilde{E}_{k}\varphi\right)\Psi_{M-1}+\frac{M\left(M-1\right)}{2}\sum_{l\in\mathbb{Z}_{\ast}^{3}}\sum_{p\in L_{l}}b_{l}^{\ast}\left(A_{l}e_{p}\right)\left[\varepsilon_{l,k}\left(e_{p};\varphi\right),b_{k}^{\ast}\left(\varphi\right)\right]\Psi_{M-2},
\end{equation}
so as $\varphi$ is an eigenvector of $2\widetilde{E}_{k}$ with eigenvalue
$\epsilon_{k}$,
\begin{equation}
\left\Vert \left(H_{\mathrm{eff}}-M\epsilon_{k}\right)\Psi_{M}\right\Vert \leq\frac{M\left(M-1\right)}{2}\left\Vert \mathcal{E} \Psi_{M-2}\right\Vert .\label{eq:SpectrumEstimate}
\end{equation}
Here the error term on the right-hand side is 
\begin{align}
\mathcal{E} & =-\frac{1}{2}\sum_{l\in\mathbb{Z}_{\ast}^{3}}\sum_{p\in L_{l}}b_{l}^{\ast}\left(A_{l}e_{p}\right)\left[\varepsilon_{l,k}\left(e_{p};\varphi\right),b_{k}^{\ast}\left(\varphi\right)\right]\\
 & =\sum_{l\in\mathbb{Z}_{\ast}^{3}}\sum_{p\in L_{k}\cap L_{l}}\sum_{q\in L_{k}}\delta_{p-l,q-k}\left\langle e_{p},\varphi\right\rangle \left\langle e_{q},\varphi\right\rangle b_{l}^{\ast}\left(A_{l}e_{p}\right)c_{q}^{\ast}c_{p-k},\nonumber 
\end{align}
where we inserted the commutator of equation (\ref{eq:ExchangeCorrectionCommutator}). This term can be rewritten as \eqref{eq:calEAlternateForm}.  

$\hfill\square$

We can now estimate the error term $\mathcal{E}$ as follows:
\begin{lem}
\label{prop:ExchangeTermEstimate}It holds that
\[
\left\Vert \mathcal{E}\Psi_{M-2}\right\Vert \leq2M\sqrt{M-1}\left\Vert \varphi\right\Vert _{\infty}^{2}\sqrt{\sum_{l\in\mathbb{Z}_{\ast}^{3}}\left\Vert \widetilde{E}_{l}-h_{l}\right\Vert _{\mathrm{HS}}^{2}}\left\Vert \Psi_{M-2}\right\Vert .
\]
\end{lem}
\textbf{Proof:} 
Write 
\begin{equation}
B_{p,q}=\sum_{l\in\mathbb{Z}_{\ast}^{3}}\delta_{p-l,q-k}1_{L_{l}}\left(p\right)b_{l}\left(A_{l}e_{p}\right)
\end{equation}
for brevity, so that $\mathcal{E}$ given by equation (\ref{eq:calEAlternateForm}) can be written as
\begin{equation}\mathcal{E}=\sum_{p,q\in L_{k}}\left\langle e_{p},\varphi\right\rangle \left\langle e_{q},\varphi\right\rangle B_{p,q}^{\ast}c_{q}^{\ast}c_{p-k}.
\end{equation}
Then $\mathcal{E}^{\ast}\mathcal{E}$ is given by
\begin{equation}
\mathcal{E}^{\ast}\mathcal{E}=\sum_{p,p',q,q'\in L_{k}}\left\langle \varphi,e_{p}\right\rangle \left\langle \varphi,e_{q}\right\rangle \left\langle e_{p'},\varphi\right\rangle \left\langle e_{q'},\varphi\right\rangle c_{p-k}^{\ast}c_{q}B_{p,q}B_{p',q'}^{\ast}c_{q'}^{\ast}c_{p'-k}.
\end{equation}
Note that by Lemma \ref{prop:ExcitationSumEstimate}, the operators $B_{p,q}^{\ast}$ obey
\begin{align}
& \sum_{p,q\in L_{k}}\left\Vert B_{p,q}^{\ast}\Psi_{M-2}\right\Vert ^{2}  \leq\sum_{p,q\in L_{k}}\sum_{l\in\mathbb{Z}_{\ast}^{3}}\delta_{p-l,q-k}1_{L_{l}}\left(p\right)\left\Vert A_{l}e_{p}\right\Vert ^{2}\left\Vert \left(\mathcal{N}_{E}+1\right)^{\frac{1}{2}}\Psi_{M-2}\right\Vert ^{2}\nonumber \\
 & =\left(M-1\right)\sum_{l\in\mathbb{Z}_{\ast}^{3}}\sum_{p\in L_{k}\cap L_{l}}\left(\sum_{q\in L_{k}}\delta_{p-l,q-k}\right)\left\Vert A_{l}e_{p}\right\Vert ^{2}\left\Vert \Psi_{M-2}\right\Vert ^{2}\label{eq:sumBpqEstimate}\\
 & \leq\left(M-1\right)\sum_{l\in\mathbb{Z}_{\ast}^{3}}\sum_{p\in L_{l}}\left\Vert A_{l}e_{p}\right\Vert ^{2}\left\Vert \Psi_{M-2}\right\Vert ^{2}=\left(M-1\right)\left(\sum_{l\in\mathbb{Z}_{\ast}^{3}}\left\Vert A_{l}\right\Vert _{\mathrm{HS}}^{2}\right)\left\Vert \Psi_{M-2}\right\Vert ^{2}.\nonumber 
\end{align}
Since $\left[c_{p-k}^{\ast}c_{q},B_{p,q}\right]=0$ it holds that
\begin{align}
 & \quad\;\;c_{p-k}^{\ast}c_{q}B_{p,q}B_{p',q'}^{\ast}c_{q'}^{\ast}c_{p'-k}=B_{p,q}c_{p-k}^{\ast}c_{q}c_{q'}^{\ast}c_{p'-k}B_{p',q'}^{\ast}\\
 & =B_{p,q}c_{q'}^{\ast}c_{p'-k}c_{p-k}^{\ast}c_{q}B_{p',q'}^{\ast}+B_{p,q}\left[c_{p-k}^{\ast}c_{q},c_{q'}^{\ast}c_{p'-k}\right]B_{p',q'}^{\ast}\nonumber 
\end{align}
so using also that $\left[c_{p-k}^{\ast}c_{q},c_{q'}^{\ast}c_{p'-k}\right]=\delta_{p,p'}\delta_{q,q'}-\delta_{p,p'}c_{q'}^{\ast}c_{q}-\delta_{q,q'}c_{p'-k}c_{p-k}^{\ast}$
we find
\begin{align}
&\left\Vert \mathcal{E}\Psi_{M-2}\right\Vert ^{2} \nonumber\\
& =\sum_{p,p',q,q'\in L_{k}}\left\langle \varphi,e_{p}\right\rangle \left\langle \varphi,e_{q}\right\rangle \left\langle e_{p'},\varphi\right\rangle \left\langle e_{q'},\varphi\right\rangle \left\langle c_{p'-k}^{\ast}c_{q'}B_{p,q}^{\ast}\Psi_{M-2},c_{p-k}^{\ast}c_{q}B_{p',q'}^{\ast}\Psi_{M-2}\right\rangle \nonumber \\
 & -\sum_{p,q,q'\in L_{k}}\left|\left\langle e_{p},\varphi\right\rangle \right|^{2}\left\langle \varphi,e_{q}\right\rangle \left\langle e_{q'},\varphi\right\rangle \left\langle c_{q'}B_{p,q}^{\ast}\Psi_{M-2},c_{q}B_{p,q'}^{\ast}\Psi_{M-2}\right\rangle \\
 & -\sum_{p,p',q\in L_{k}}\left|\left\langle e_{q},\varphi\right\rangle \right|^{2}\left\langle \varphi,e_{p}\right\rangle \left\langle e_{p'},\varphi\right\rangle \left\langle c_{p'-k}^{\ast}B_{p,q}^{\ast}\Psi_{M-2},c_{p-k}^{\ast}B_{p',q}^{\ast}\Psi_{M-2}\right\rangle \nonumber \\
 & +\sum_{p,q\in L_{k}}\left|\left\langle e_{p},\varphi\right\rangle \right|^{2}\left|\left\langle e_{q},\varphi\right\rangle \right|^{2}\left\Vert B_{p,q}^{\ast}\Psi_{M-2}\right\Vert ^{2}=:T_{1}+T_{2}+T_{3}+T_{4}.\nonumber 
\end{align}
We estimate the separate terms. For $T_{1}$ we can apply the Cauchy-Schwarz
inequality and equation (\ref{eq:sumBpqEstimate}) to bound
\begin{align}
\left|T_{1}\right| & \leq\left\Vert \varphi\right\Vert _{\infty}^{4}\sum_{p,p',q,q'\in L_{k}}\left\Vert c_{p'-k}^{\ast}c_{q'}B_{p,q}^{\ast}\Psi_{M-2}\right\Vert \left\Vert c_{p-k}^{\ast}c_{q}B_{p',q'}^{\ast}\Psi_{M-2}\right\Vert \nonumber \\
 & \leq\left\Vert \varphi\right\Vert _{\infty}^{4}\sum_{p,p',q,q'\in L_{k}}\left\Vert c_{p-k}^{\ast}c_{q}B_{p',q'}^{\ast}\Psi_{M-2}\right\Vert ^{2}\leq\left\Vert \varphi\right\Vert _{\infty}^{4}\sum_{p',q'\in L_{k}}\left\Vert \mathcal{N}_{E}B_{p',q'}^{\ast}\Psi_{M-2}\right\Vert ^{2}\\
 & \leq\left(M-1\right)^{2}\left\Vert \varphi\right\Vert _{\infty}^{4}\sum_{p',q'\in L_{k}}\left\Vert B_{p',q'}^{\ast}\Psi_{M-2}\right\Vert ^{2}\leq\left(M-1\right)^{3}\left\Vert \varphi\right\Vert _{\infty}^{4}\left(\sum_{l\in\mathbb{Z}_{\ast}^{3}}\left\Vert A_{l}\right\Vert _{\mathrm{HS}}^{2}\right)\left\Vert \Psi_{M-2}\right\Vert ^{2},\nonumber 
\end{align}
and $T_{2}$ is similarly bounded as
\begin{align}
\left|T_{2}\right| & \leq\left\Vert \varphi\right\Vert _{\infty}^{4}\sum_{p,q,q'\in L_{k}}\left\Vert c_{q'}B_{p,q}^{\ast}\Psi_{M-2}\right\Vert \left\Vert c_{q}B_{p,q'}^{\ast}\Psi_{M-2}\right\Vert \nonumber \\
 & \leq\left\Vert \varphi\right\Vert _{\infty}^{4}\sum_{p\in L_{k}}\sqrt{\sum_{q,q'\in L_{k}}\left\Vert c_{q'}B_{p,q}^{\ast}\Psi_{M-2}\right\Vert ^{2}}\sqrt{\sum_{q,q'\in L_{k}}\left\Vert c_{q}B_{p,q'}^{\ast}\Psi_{M-2}\right\Vert ^{2}}\\
 & \leq\left\Vert \varphi\right\Vert _{\infty}^{4}\sum_{p,q\in L_{k}}\left\Vert \mathcal{N}_{E}^{\frac{1}{2}}B_{p,q}^{\ast}\Psi_{M-2}\right\Vert ^{2}\leq\left(M-1\right)^{2}\left\Vert \varphi\right\Vert _{\infty}^{4}\left(\sum_{l\in\mathbb{Z}_{\ast}^{3}}\left\Vert A_{l}\right\Vert _{\mathrm{HS}}^{2}\right)\left\Vert \Psi_{M-2}\right\Vert ^{2},\nonumber 
\end{align}
the same estimate holding also for $T_{3}$. Finally $T_{4}$ is just
bounded by
\begin{equation}
\left|T_{4}\right|\leq\left(M-1\right)^{2}\left\Vert \varphi\right\Vert _{\infty}^{4}\left(\sum_{l\in\mathbb{Z}_{\ast}^{3}}\left\Vert A_{l}\right\Vert _{\mathrm{HS}}^{2}\right)\left\Vert \Psi_{M-2}\right\Vert ^{2}
\end{equation}
so combining the estimates we find
\begin{align}
\left\Vert \mathcal{E}\Psi_{M-2}\right\Vert  & \leq M\sqrt{M-1}\left\Vert \varphi\right\Vert _{\infty}^{2}\sqrt{\sum_{l\in\mathbb{Z}_{\ast}^{3}}\left\Vert A_{l}\right\Vert _{\mathrm{HS}}^{2}}\left\Vert \Psi_{M-2}\right\Vert \\
 & =2M\sqrt{M-1}\left\Vert \varphi\right\Vert _{\infty}^{2}\sqrt{\sum_{l\in\mathbb{Z}_{\ast}^{3}}\left\Vert \widetilde{E}_{l}-h_{l}\right\Vert _{\mathrm{HS}}^{2}}\left\Vert \Psi_{M-2}\right\Vert .\nonumber 
\end{align}
$\hfill\square$

Proposition \ref{prop:SpectralEstimate} now follows by combining Lemma \ref{prop:4.23} with Lemma  \ref{prop:ExchangeTermEstimate} and Proposition \ref{prop:NormLowerBound} to see that
\begin{align}
 & \quad\left\Vert \left(H_{\mathrm{eff}}-M\epsilon_{k}\right)\hat{\Psi}_{M}\right\Vert \leq M\left(M-1\right)\frac{\left\Vert \mathcal{E}\Psi_{M-2}\right\Vert }{\left\Vert \Psi_{M}\right\Vert }\nonumber \\
 & \leq2M^2\left(M-1\right) ^{\frac{3}{2}}\left\Vert \varphi\right\Vert _{\infty}^{2}\sqrt{\sum_{l\in\mathbb{Z}_{\ast}^{3}}\left\Vert \widetilde{E}_{l}-h_{l}\right\Vert _{\mathrm{HS}}^{2}}\frac{\left\Vert \Psi_{M-2}\right\Vert }{\left\Vert \Psi_{M}\right\Vert }\nonumber \\
 & \leq 2M^2\left(M-1\right) ^{\frac{3}{2}} \left\Vert \varphi\right\Vert _{\infty}^{2}\sqrt{\sum_{l\in\mathbb{Z}_{\ast}^{3}}\left\Vert \widetilde{E}_{l}-h_{l}\right\Vert _{\mathrm{HS}}^{2}}\sqrt{\frac{\left(M-2\right)!}{M!\left(1-\frac{M\left(M-1\right)}{2}\left\Vert \varphi\right\Vert _{6}^{3}\right)}}\label{eq:SpectrumEstimate2}\\
 & =2M^{\frac{3}{2}} (M-1) \left\Vert \varphi\right\Vert _{\infty}^{2}\sqrt{\sum_{l\in\mathbb{Z}_{\ast}^{3}}\left\Vert \widetilde{E}_{l}-h_{l}\right\Vert _{\mathrm{HS}}^{2}}\frac{1}{\sqrt{1-\frac{M\left(M-1\right)}{2}\left\Vert \varphi\right\Vert _{6}^{3}}}\nonumber \\
 & \leq\frac{2\left\Vert \varphi\right\Vert _{\infty}^{2}\sqrt{\sum_{l\in\mathbb{Z}_{\ast}^{3}}\left\Vert \widetilde{E}_{l}-h_{l}\right\Vert _{\mathrm{HS}}^{2}}}{\sqrt{1-M^{2}\left\Vert \varphi\right\Vert _{6}^{3}}}M^{\frac{5}{2}}.\nonumber 
\end{align}

\section{Estimation of One-Body Quantities} \label{sec:EstimationofOne-BodyQuantities}

To proceed we must now derive some estimates on the one-body quantities
involved - we need to verify that $\varphi$ is indeed ``almost delocalized''
and bound $\sum_{l\in\mathbb{Z}_{\ast}^{3}}\left\Vert \widetilde{E}_{l}-h_{l}\right\Vert _{\mathrm{HS}}^{2}$.
We prove the following:
\begin{prop}
\label{prop:ExplicitOneBodyQuantities} For $\left|k\right|\ll\sqrt{k_{F}}$, it holds that
\[
\left\Vert \varphi\right\Vert _{\infty}^{2},\,\left\Vert \varphi\right\Vert _{6}^{3}\leq \frac{C}{k_{F}^{2}\left|k\right|}, 
\]
and
\[
\sum_{l\in\mathbb{Z}_{\ast}^{3}}\left\Vert \widetilde{E}_{l}-h_{l}\right\Vert _{\mathrm{HS}}^{2}\leq Ck_{F}^{5}
\]
for $C>0$ independent of $k$ and $k_{F}$.
\end{prop}
Note that the condition $\left|k\right| \ll\sqrt{k_{F}}$, namely $\left|k\right|/ \sqrt{k_{F}} \to 0$ as $k_F\to \infty$, holds if $|k| \le k_F^{\delta}$ with $\delta<1/2$. 

\subsection{General Estimates} 

To avoid unnecessary subscripts we consider instead of $\ell^{2}\left(L_{k}\right)$
a general $n$-dimensional inner product space $\left(V,\left\langle \cdot,\cdot\right\rangle \right)$,
on which a positive symmetric operator $h:V\rightarrow V$ acts, with
diagonalizing basis $\left(e_{i}\right)_{i=1}^{n}$ and eigenvalues
$\left(\lambda_{i}\right)_{i=1}^{n}$, and a fixed $v\in V$ such
that $\left\langle e_{i},v\right\rangle >0$ for all $1\leq i\leq n$.

Define $\widetilde{E}:V\rightarrow V$ by
\begin{equation}
\tilde{E}=\left(h^{\frac{1}{2}}\left(h+2P_{v}\right)h^{\frac{1}{2}}\right)^{\frac{1}{2}}=\left(h^{2}+2P_{h^{\frac{1}{2}}v}\right)^{\frac{1}{2}}.
\end{equation}
Let $\varphi\in V$ be a normalized eigenvector of $\widetilde{E}$
with greatest eigenvalue $\epsilon$ ($>\left\Vert h\right\Vert $)
(note that we do not include the factor of $2$ in this section),
with phase chosen such that $\left\langle h^{\frac{1}{2}}v,\varphi\right\rangle \geq0$.
Then as $\widetilde{E}$ squares to $h^{2}+2P_{h^{\frac{1}{2}}v}$,
we have
\begin{equation}
\epsilon^{2}\varphi=h^{2}\varphi+2\left\langle h^{\frac{1}{2}}v,\varphi\right\rangle h^{\frac{1}{2}}v
\end{equation}
whence
\begin{equation}
\varphi=2\left\langle h^{\frac{1}{2}}v,\varphi\right\rangle \left(\epsilon^{2}-h^{2}\right)^{-1}h^{\frac{1}{2}}v.
\end{equation}
This identity lets us describe the components of $\varphi$ (with
respect to $\left(e_{i}\right)_{i=1}^{n}$) in terms of the single
unknown $\epsilon$: Taking the inner product with $e_{i}$ yields
\begin{equation}
\left\langle e_{i},\varphi\right\rangle =2\left\langle h^{\frac{1}{2}}v,\varphi\right\rangle \left\langle e_{i},\left(\epsilon^{2}-h^{2}\right)^{-1}h^{\frac{1}{2}}v\right\rangle =2\left\langle h^{\frac{1}{2}}v,\varphi\right\rangle \frac{\sqrt{\lambda_{i}}}{\epsilon^{2}-\lambda_{i}^{2}}\left\langle e_{i},v\right\rangle 
\end{equation}
and now we may note that $2\left\langle h^{\frac{1}{2}}v,\varphi\right\rangle $
is simply a constant independent of $i$. As $\varphi$ is by assumption
normalized, we thus have
\begin{equation}
\left\langle e_{i},\varphi\right\rangle =\frac{1}{\sqrt{\sum_{i=1}^{N}\frac{\lambda_{i}}{\left(\epsilon^{2}-\lambda_{i}^{2}\right)^{2}}\left|\left\langle e_{i},v\right\rangle \right|^{2}}}\frac{\sqrt{\lambda_{i}}}{\epsilon^{2}-\lambda_{i}^{2}}\left\langle e_{i},v\right\rangle ,\quad1\leq i\leq n.
\end{equation}
Note that by the variational principle $\epsilon^{2} \ge \tilde{E}^2 =  h^{2}+2P_{h^{\frac{1}{2}}v}$, we have
\begin{equation} \label{eq:5.6}
\epsilon^{2}\geq\frac{\left\langle h^{\frac{1}{2}}v,\left(h^{2}+2P_{h^{\frac{1}{2}}v}\right)h^{\frac{1}{2}}v\right\rangle }{\left\langle h^{\frac{1}{2}}v,h^{\frac{1}{2}}v\right\rangle }=2\left\langle v,hv\right\rangle +\frac{\left\langle v,h^{3}v\right\rangle }{\left\langle v,hv\right\rangle } \ge 2 \left\langle v,hv\right\rangle .
\end{equation}
So we immediately obtain the following:
\begin{lem}
\label{prop:phiComponentEstimate}Let $\lambda_{\max}=\max_{1\leq i\leq n}\lambda_{i}$.
Then provided $2\left\langle v,hv\right\rangle >\lambda_{\max}^{2}$
it holds that
\[
\left|\left\langle e_{i},\varphi\right\rangle \right|\leq\frac{2\left\langle v,hv\right\rangle }{2\left\langle v,hv\right\rangle -\lambda_{\max}^{2}}\frac{\sqrt{\lambda_{i}}}{\sqrt{\left\langle v,hv\right\rangle }}\left|\left\langle e_{i},v\right\rangle \right|,\quad1\leq i\leq n.
\]
\end{lem}
\textbf{Proof:} We simply estimate
\begin{align}
\left|\left\langle e_{i},\varphi\right\rangle \right| & =\frac{1}{\sqrt{\sum_{i=1}^{N}\frac{\lambda_{i}}{\left(\epsilon^{2}-\lambda_{i}^{2}\right)^{2}}\left|\left\langle e_{i},v\right\rangle \right|^{2}}}\frac{\sqrt{\lambda_{i}}}{\epsilon^{2}-\lambda_{i}^{2}}\left|\left\langle e_{i},v\right\rangle \right|\nonumber \\
 & \leq\frac{1}{\frac{1}{\epsilon^{2}}\sqrt{\sum_{i=1}^{N}\lambda_{i}\left|\left\langle e_{i},v\right\rangle \right|^{2}}}\frac{1}{\epsilon^{2}-\lambda_{\max}^{2}}\sqrt{\lambda_{i}}\left|\left\langle e_{i},v\right\rangle \right|\\
 & =\frac{\epsilon^{2}}{\epsilon^{2}-\lambda_{\max}^{2}}\frac{\sqrt{\lambda_{i}}}{\left\langle v,hv\right\rangle }\left|\left\langle e_{i},v\right\rangle \right|\nonumber 
\end{align}
and note that by \eqref{eq:5.6}, 
\begin{equation}
\frac{\epsilon^{2}}{\epsilon^{2}-\lambda_{\max}^{2}}=\frac{1}{1-\frac{\lambda_{\max}^{2}}{\epsilon^{2}}}\leq\frac{1}{1-\frac{\lambda_{\max}^{2}}{2\left\langle v,hv\right\rangle }}=\frac{2\left\langle v,hv\right\rangle }{2\left\langle v,hv\right\rangle -\lambda_{\max}^{2}}.
\end{equation}
$\hfill\square$

For the statement of Theorem \ref{thm:CoulombPlasmonStates}
it is also interesting to bound $\epsilon$ from above: We just saw
that
\begin{equation}
\epsilon^{2}\geq2\left\langle v,hv\right\rangle +\frac{\left\langle v,h^{3}v\right\rangle }{\left\langle v,hv\right\rangle }
\end{equation}
and the right-hand side is in fact the leading contribution to $\epsilon^{2}$:
\begin{lem}
\label{prop:epskBound}Provided $2\left\langle v,hv\right\rangle >\lambda_{\max}^{2}$
it holds that
\[
\epsilon^{2}\leq2\left\langle v,hv\right\rangle +\frac{\left\langle v,h^{3}v\right\rangle }{\left\langle v,hv\right\rangle }+\frac{4\left\langle v,h^{3}v\right\rangle \lambda_{\max}^{2}}{\left(2\left\langle v,hv\right\rangle -\lambda_{\max}^{2}\right)^{2}}.
\]
\end{lem}
\textbf{Proof:} By the identity $\varphi=\left\Vert \left(\epsilon^{2}-h^{2}\right)^{-1}h^{\frac{1}{2}}v\right\Vert ^{-1}\left(\epsilon^{2}-h^{2}\right)^{-1}h^{\frac{1}{2}}v$
we have that
\begin{align}
\left\langle \varphi,h^{2}\varphi\right\rangle  & =\frac{\left\langle v,h^{3}\left(\epsilon^{2}-h^{2}\right)^{-2}v\right\rangle }{\left\langle v,h\left(\epsilon^{2}-h^{2}\right)^{-2}v\right\rangle }\leq\frac{\frac{1}{\left(\epsilon^{2}-\lambda_{\max}^{2}\right)^{2}}\left\langle v,h^{3}v\right\rangle }{\frac{1}{\epsilon^{4}}\left\langle v,hv\right\rangle }\\
 & =\frac{\epsilon^{4}}{\left(\epsilon^{2}-\lambda_{\max}^{2}\right)^{2}}\frac{\left\langle v,h^{3}v\right\rangle }{\left\langle v,hv\right\rangle }\leq\frac{\left(2\left\langle v,hv\right\rangle \right)^{2}}{\left(2\left\langle v,hv\right\rangle -\lambda_{\max}^{2}\right)^{2}}\frac{\left\langle v,h^{3}v\right\rangle }{\left\langle v,hv\right\rangle }\nonumber 
\end{align}
where we estimated as above. Continuing the estimate we then find
\begin{align}
\left\langle \varphi,h^{2}\varphi\right\rangle  & \leq\frac{\left\langle v,h^{3}v\right\rangle }{\left\langle v,hv\right\rangle }+\frac{\left(2\left\langle v,hv\right\rangle \right)^{2}-\left(2\left\langle v,hv\right\rangle -\lambda_{\max}^{2}\right)^{2}}{\left(2\left\langle v,hv\right\rangle -\lambda_{\max}^{2}\right)^{2}}\frac{\left\langle v,h^{3}v\right\rangle }{\left\langle v,hv\right\rangle }\\
 & =\frac{\left\langle v,h^{3}v\right\rangle }{\left\langle v,hv\right\rangle }+\frac{4\left\langle v,hv\right\rangle \lambda_{\max}^{2}-\lambda_{\max}^{4}}{\left(2\left\langle v,hv\right\rangle -\lambda_{\max}^{2}\right)^{2}}\frac{\left\langle v,h^{3}v\right\rangle }{\left\langle v,hv\right\rangle }\leq\frac{\left\langle v,h^{3}v\right\rangle }{\left\langle v,hv\right\rangle }+\frac{4\left\langle v,h^{3}v\right\rangle \lambda_{\max}^{2}}{\left(2\left\langle v,hv\right\rangle -\lambda_{\max}^{2}\right)^{2}}.\nonumber 
\end{align}
From the eigenvalue equation for $\epsilon^{2}$ we can then conclude
that
\begin{align}
\epsilon^{2} & =\left\langle \varphi,h^{2}\varphi\right\rangle +2\left|\left\langle h^{\frac{1}{2}}v,\varphi\right\rangle \right|^{2}\leq2\left\Vert h^{\frac{1}{2}}v\right\Vert ^{2}\left\Vert \varphi\right\Vert ^{2}+\frac{\left\langle v,h^{3}v\right\rangle }{\left\langle v,hv\right\rangle }+\frac{4\left\langle v,h^{3}v\right\rangle \lambda_{\max}^{2}}{\left(2\left\langle v,hv\right\rangle -\lambda_{\max}^{2}\right)^{2}} \nonumber \\
 & =2\left\langle v,hv\right\rangle +\frac{\left\langle v,h^{3}v\right\rangle }{\left\langle v,hv\right\rangle }+\frac{4\left\langle v,h^{3}v\right\rangle \lambda_{\max}^{2}}{\left(2\left\langle v,hv\right\rangle -\lambda_{\max}^{2}\right)^{2}}.
\end{align}
$\hfill\square$

\subsection*{Estimating $\left\Vert \widetilde{E}-h\right\Vert _{\mathrm{HS}}$}

Finally we consider $\left\Vert \widetilde{E}-h\right\Vert _{\mathrm{HS}}=\left\Vert \left(h^{2}+2P_{h^{\frac{1}{2}}v}\right)^{\frac{1}{2}}-h\right\Vert _{\mathrm{HS}}$.
In \cite[Eq. (7.22)]{CHN-21} we derived the identity
\begin{equation}
\widetilde{E}-h=\frac{4}{\pi}\int_{0}^{\infty}\frac{t^{2}}{1+2\left\langle v,h\left(h^{2}+t^{2}\right)^{-1}v\right\rangle }P_{\left(h^{2}+t^{2}\right)^{-1}h^{\frac{1}{2}}v}dt
\end{equation}
from which it follows that
\begin{equation}
0\leq\left\langle e_{i},\left(\widetilde{E}-h\right)e_{j}\right\rangle \leq2\frac{\sqrt{\lambda_{i}\lambda_{j}}}{\lambda_{i}+\lambda_{j}}\left\langle e_{i},v\right\rangle \left\langle v,e_{j}\right\rangle ,\quad1\leq i,j\leq n.
\end{equation}
This is asymptotically optimal for ``small $v$'', but without the
mean-field scaling we also need to consider ``large $v$''. While
a direct elementwise estimate appears to be more involved in this
regime, a good Hilbert-Schmidt estimate is in fact simpler. Covering
both regimes, we have the following:
\begin{lem}
\label{prop:ExchangeHilbertSchmidtEstimate}It holds that
\[
\left\Vert \widetilde{E}-h\right\Vert _{\mathrm{HS}}^{2}\leq\min\left\{ 2\left\langle v,hv\right\rangle ,4\left\Vert v\right\Vert ^{4}\right\} .
\]
\end{lem}
\textbf{Proof:} We first note that
\begin{align}
\left\Vert \widetilde{E}-h\right\Vert _{\mathrm{HS}}^{2} & =\mathrm{tr}\left(\left(\widetilde{E}-h\right)^{2}\right)=\mathrm{tr}\left(\widetilde{E}^{2}+h^{2}-\widetilde{E}h-h\widetilde{E}\right)\\
 & =2\,\mathrm{tr}\left(h^{2}+P_{h^{\frac{1}{2}}v}-h^{\frac{1}{2}}\widetilde{E}h^{\frac{1}{2}}\right)\nonumber 
\end{align}
as $\widetilde{E}^{2}=h^{2}+2P_{h^{\frac{1}{2}}v}$. Since we may
trivially estimate that $\widetilde{E}\geq h$ we can then conclude
\begin{equation}
\left\Vert \widetilde{E}-h\right\Vert _{\mathrm{HS}}^{2}\leq2\,\mathrm{tr}\left(h^{2}+P_{h^{\frac{1}{2}}v}-h^{2}\right)=2\left\Vert h^{\frac{1}{2}}v\right\Vert ^{2}=2\left\langle v,hv\right\rangle .
\end{equation}
For the other estimate we simply apply the elementwise estimate:
\begin{align}
\left\Vert \widetilde{E}-h\right\Vert _{\mathrm{HS}}^{2} & =\sum_{i,j=1}^{n}\left|\left\langle e_{i},\left(\tilde{E}-h\right)e_{j}\right\rangle \right|^{2}\leq4\sum_{i,j=1}^{n}\frac{\lambda_{i}\lambda_{j}}{\left(\lambda_{i}+\lambda_{j}\right)^{2}}\left|\left\langle e_{i},v\right\rangle \left\langle v,e_{j}\right\rangle \right|^{2}\\
 & \leq4\sum_{i,j=1}^{n}\frac{\lambda_{i}\lambda_{j}}{\left(\lambda_{i}+\lambda_{j}\right)^{2}}\left|\left\langle e_{i},v\right\rangle \left\langle v,e_{j}\right\rangle \right|^{2}\leq4\left(\sum_{i=1}^{n}\left|\left\langle e_{i},v\right\rangle \right|^{2}\right)^{2}=4\left\Vert v\right\Vert ^{4}.\nonumber 
\end{align}
$\hfill\square$

\subsection{Proof of Proposition \ref{prop:ExplicitOneBodyQuantities}}


To prove Proposition \ref{prop:ExplicitOneBodyQuantities} we now
only need to insert the specific one-body operators of our problem; recall
that in this case
\begin{equation}
h_{k}e_{p}=\lambda_{k,p}e_{p},\quad\lambda_{k,p}=\frac{1}{2}\left(\left|p\right|^{2}-\left|p-k\right|^{2}\right)
\end{equation}
and
\begin{equation}
v_{k}=\sqrt{\frac{\hat{V}_{k}}{2\left(2\pi\right)^{3}}}\sum_{p\in L_{k}}e_{p}.
\end{equation}
First, for $\left\Vert \varphi\right\Vert _{6}^{3}$ and $\left\Vert \varphi\right\Vert _{\infty}^{2}$,
we trivially have that
\begin{equation} \label{eq:varphi-6-0}
\left\Vert \varphi\right\Vert _{6}^{3}=\sqrt{\sum_{p\in L_{k}}\left|\left\langle e_{p},\varphi\right\rangle \right|^{6}}\leq\sqrt{\left|L_{k}\right|}\left\Vert \varphi\right\Vert _{\infty}^{3}
\end{equation}
and by Lemma \ref{prop:phiComponentEstimate} we have the estimate
\begin{equation} \label{eq:varphi-inf-0}
\left\Vert \varphi\right\Vert _{\infty}\leq\frac{2\left\langle v_{k},h_{k}v_{k}\right\rangle }{2\left\langle v_{k},h_{k}v_{k}\right\rangle -\lambda_{k,\max}^{2}}\frac{\sqrt{\lambda_{k,\max}}}{\sqrt{\left\langle v_{k},h_{k}v_{k}\right\rangle }}\max_{p\in L_{k}}\left|\left\langle e_{p},v_{k}\right\rangle \right|
\end{equation}
provided $2\left\langle v_{k},h_{k}v_{k}\right\rangle -\lambda_{k,\max}^{2}>0$. Moreover, under the condition that $\left|k\right|\ll\sqrt{k_{F}}$, the $k$-dependent quantities
behave as (recall that $\hat{V}_{k}=g\left|k\right|^{-2}$)
\begin{equation} \label{eq:need-to-mod-1}
\lambda_{k,\max}\sim k_{F}\left|k\right|,\quad\max_{p\in L_{k}}\left|\left\langle e_{p},v_{k}\right\rangle \right|=\sqrt{\frac{\hat{V}_{k}}{2\left(2\pi\right)^{3}}}\sim\sqrt{\hat{V}_{k}}\sim\left|k\right|^{-1}
\end{equation}
and
\begin{equation}\label{eq:need-to-mod-2}
\left\langle v_{k},h_{k}v_{k}\right\rangle =\frac{\hat{V}_{k}}{2\left(2\pi\right)^{3}}\sum_{p\in L_{k}}\lambda_{k,p}\sim k_{F}^{3}.
\end{equation}
Here note that the behavior $\lambda_{k,\max}\sim k_{F}\left|k\right|$ can be deduced easily from \eqref{eq:2.45}. So the estimate on $\left\Vert \varphi\right\Vert _{\infty}$ in \eqref{eq:varphi-inf-0} boils down to 
\begin{equation}
\left\Vert \varphi\right\Vert _{\infty}\leq C\frac{k_{F}^{3}}{k_{F}^{3}-C'k_{F}^{2}\left|k\right|^{2}}\frac{\sqrt{k_{F}\left|k\right|}}{\sqrt{k_{F}^{3}}}\left|k\right|^{-1}\leq C\frac{1}{\sqrt{k_{F}^{2}\left|k\right|}}. 
\end{equation}
The estimate on $\left\Vert \varphi\right\Vert _{6}$ in \eqref{eq:varphi-6-0} can be simplified using $\sqrt{\left|L_{k}\right|}\sim\sqrt{k_{F}^{2}\left|k\right|}$. 
It follows that
\begin{equation}
\left\Vert \varphi\right\Vert _{\infty}^{2},\left\Vert \varphi\right\Vert _{6}^{3}\leq \frac{C}{k_{F}^{2}\left|k\right|}
\end{equation}
when $\left|k\right|\ll\sqrt{k_{F}}$, as claimed.

For $\sum_{l\in\mathbb{Z}_{\ast}^{3}}\left\Vert \widetilde{E}_{l}-h\right\Vert _{\mathrm{HS}}^{2}$
we note that by Lemma \ref{prop:ExchangeHilbertSchmidtEstimate},
when $\left|l\right|\leq2k_{F}$,
\begin{align}
\left\Vert \widetilde{E}_{l}-h_{l}\right\Vert _{\mathrm{HS}}^{2} & \leq\min\left\{ 2\left\langle v_{l},h_{l}v_{l}\right\rangle ,4\left\Vert v_{l}\right\Vert ^{4}\right\} \leq C\min\left\{ k_{F}^{3},k_{F}^{4}\left|l\right|^{-2}\right\} \\
 & =Ck_{F}^{3}\min\left\{ 1,k_{F}\left|l\right|^{-2}\right\} \nonumber 
\end{align}
and when $\left|l\right|>2k_{F}$
\begin{equation}
\left\Vert \widetilde{E}_{l}-h_{l}\right\Vert _{\mathrm{HS}}^{2}\leq4\left\Vert v_{l}\right\Vert ^{4}\leq C\hat{V}_{k}^{2}\left|B_{F}\right|^{2}\leq Ck_{F}^{6}\left|l\right|^{-4},
\end{equation}
whence
\begin{align}
\sum_{l\in\mathbb{Z}_{\ast}^{3}}\left\Vert \widetilde{E}_{l}-h\right\Vert _{\mathrm{HS}}^{2} & \leq C\sum_{l\in B\left(0,\sqrt{k_{F}}\right)\cap\mathbb{Z}^{3}}k_{F}^{3}+C\sum_{l\in\overline{B}\left(0,2k_{F}\right)\backslash B\left(0,\sqrt{k_{F}}\right)\cap\mathbb{Z}^{3}}k_{F}^{4}\left|l\right|^{-2}\nonumber \\
 & +C\sum_{l\in\mathbb{Z}^{3}\backslash\overline{B}\left(0,2k_{F}\right)}k_{F}^{6}\left|l\right|^{-4}\\
 & \leq Ck_{F}^{4+\frac{1}{2}}+Ck_{F}^{5}+Ck_{F}^{5}\leq Ck_{F}^{5}\nonumber 
\end{align}
as claimed. 

$\hfill\square$

\section{Conclusion} \label{sec:Conclusion}

We can now conclude the proof of the main result. 

\medskip

\textbf{Proof of Theorem \ref{thm:CoulombPlasmonStates}:} For the first part of Theorem \ref{thm:CoulombPlasmonStates}, by Proposition \ref{prop:SpectralEstimate} and the
estimates of Proposition \ref{prop:ExplicitOneBodyQuantities} we
have
\begin{align}
\left\Vert \left(H_{\mathrm{eff}}-M\epsilon_{k}\right)\hat{\Psi}_{M}\right\Vert  & \leq C\frac{1}{k_{F}^{2}\left|k\right|}\frac{\sqrt{\sum_{l\in\mathbb{Z}_{\ast}^{3}}\left\Vert \widetilde{E}_{l}-h_{l}\right\Vert _{\mathrm{HS}}^{2}}}{\sqrt{1-\frac{M^{2}}{k_{F}^{2}\left|k\right|}}}M^{\frac{5}{2}}\nonumber \\
 & \leq C\left|k\right|^{-1}\sqrt{k_{F}^{-4}\sum_{l\in\mathbb{Z}_{\ast}^{3}}\left\Vert \widetilde{E}_{l}-h_{l}\right\Vert _{\mathrm{HS}}^{2}}M^{\frac{5}{2}}\\
 & \leq C\left|k\right|^{-1}\sqrt{k_{F}}M^{\frac{5}{2}}\nonumber 
\end{align}
where the assumption $\left|k\right|\leq k_{F}^{\delta}$ ensures
the applicability of Proposition \ref{prop:ExplicitOneBodyQuantities}
and the condition $M\leq k_{F}^{\varepsilon}$ ensures that $\left(1-\frac{M^{2}}{k_{F}^{2}\left|k\right|}\right)^{-\frac{1}{2}}\leq C$.

For the second part of Theorem \ref{thm:CoulombPlasmonStates}, concerning $\epsilon_{k}$,
we have by Lemma \ref{prop:epskBound} that (remembering to
include a factor of $2$)
\begin{equation}
\epsilon_{k}\leq\sqrt{8\left\langle v_{k},h_{k}v_{k}\right\rangle +4\frac{\left\langle v_{k},h_{k}^{3}v_{k}\right\rangle }{\left\langle v_{k},h_{k}v_{k}\right\rangle }+\frac{16\left\langle v_{k},h_{k}^{3}v_{k}\right\rangle \lambda_{k,\max}^{2}}{\left(2\left\langle v_{k},h_{k}v_{k}\right\rangle -\lambda_{k,\max}^{2}\right)^{2}}}
\end{equation}
and
\begin{equation}
\epsilon_{k}\geq\sqrt{8\left\langle v_{k},h_{k}v_{k}\right\rangle +4\frac{\left\langle v_{k},h_{k}^{3}v_{k}\right\rangle }{\left\langle v_{k},h_{k}v_{k}\right\rangle }}\geq C\sqrt{\left\langle v_{k},h_{k}v_{k}\right\rangle }\geq Ck_{F}^{\frac{3}{2}}.
\end{equation}
As $\sqrt{a+b}-\sqrt{a}\leq\frac{b}{2\sqrt{a}}$ we may then estimate
\begin{align}
 & \epsilon_{k}-\sqrt{8\left\langle v_{k},h_{k}v_{k}\right\rangle +4\frac{\left\langle v_{k},h_{k}^{3}v_{k}\right\rangle }{\left\langle v_{k},h_{k}v_{k}\right\rangle }}\nonumber \\
 & \leq\frac{1}{2\sqrt{8\left\langle v_{k},h_{k}v_{k}\right\rangle +4\frac{\left\langle v_{k},h_{k}^{3}v_{k}\right\rangle }{\left\langle v_{k},h_{k}v_{k}\right\rangle }}}\frac{16\left\langle v_{k},h_{k}^{3}v_{k}\right\rangle \lambda_{k,\max}^{2}}{\left(2\left\langle v_{k},h_{k}v_{k}\right\rangle -\lambda_{k,\max}^{2}\right)^{2}}\\
 & \leq C\frac{1}{k_{F}^{\frac{3}{2}}}\frac{k_{F}^{5}\left|k\right|^{2}\left(k_{F}\left|k\right|\right)^{2}}{k_{F}^{6}}=Ck_{F}^{-\frac{1}{2}}\left|k\right|^{4}\nonumber 
\end{align}
for the claim that
\begin{equation}
\epsilon_{k}=\sqrt{8\left\langle v_{k},h_{k}v_{k}\right\rangle +4\frac{\left\langle v_{k},h_{k}^{3}v_{k}\right\rangle }{\left\langle v_{k},h_{k}v_{k}\right\rangle }}+O\left(k_{F}^{-\frac{1}{2}}\left|k\right|^{4}\right).
\end{equation}
The proof of Theorem \ref{thm:CoulombPlasmonStates} is complete. 

$\hfill\square$

\subsection*{Further explanation of \eqref{eq:eps-k-af-thm} for $\epsilon_{k}$ in the Thermodynamic Limit}

In the thermodynamic limit, in which we replace Riemann sums by the
corresponding integrals, we have
\begin{equation}
\left\langle v_{k},h_{k}^{\beta}v_{k}\right\rangle =\frac{\hat{V}_{k}}{2\left(2\pi\right)^{3}}\sum_{p\in L_{k}}\lambda_{k,p}^{\beta}\sim\frac{\hat{V}_{k}}{2\left(2\pi\right)^{3}}\int_{\mathcal{L}_{k}}\left(k\cdot p-\frac{1}{2}\left|k\right|^{2}\right)^{\beta}dp
\end{equation}
where $\mathcal{L}_{k}=\left\{ p\in\mathbb{R}^{3}\mid\left|p-k\right|\leq k_{F}<\left|p\right|\right\} $
is now the ``solid'' lune. By integrating along the $k\cdot p=\text{constant}$
planes one may reexpress the integral, when $\left|k\right|\leq2k_{F}$,
as
\begin{align}
\int_{\mathcal{L}_{k}}f\left(k\cdot p-\frac{1}{2}\left|k\right|^{2}\right)dp & =2\pi\left|k\right|\int_{\frac{1}{2}\left|k\right|}^{k_{F}}f\left(\left|k\right|\left(t-\frac{1}{2}\left|k\right|\right)\right)\left(t-\frac{1}{2}\left|k\right|\right)dt\\
 & \quad +\pi\int_{k_{F}}^{k_{F}+\left|k\right|}f\left(\left|k\right|\left(t-\frac{1}{2}\left|k\right|\right)\right)\left(k_{F}^{2}-\left(t-\left|k\right|\right)^{2}\right)dt\nonumber 
\end{align}
with $f(x) = x^\beta.$ It follows that for Coulomb
\begin{align}
\left\langle v_{k},h_{k}v_{k}\right\rangle  & \sim\frac{\hat{V}_{k}}{2\left(2\pi\right)^{3}}\left(\frac{2\pi}{3}k_{F}^{3}\left|k\right|^{2}\right)=\frac{1}{4}\frac{g}{6\pi^{2}}k_{F}^{3}, \\
\left\langle v_{k},h_{k}^{3}v_{k}\right\rangle  & \sim\frac{\hat{V}_{k}}{2\left(2\pi\right)^{3}}\left(\frac{2\pi}{5}k_{F}^{5}\left|k\right|^{4}+\frac{\pi}{6}k_{F}^{3}\left|k\right|^{6}\right)\approx\frac{1}{4}\frac{g}{10\pi^{2}}k_{F}^{5}\left|k\right|^{2}\nonumber 
\end{align}
whence
\begin{equation}
\epsilon_{k}\approx\sqrt{\frac{g}{3\pi^{2}}k_{F}^{3}+4\frac{\frac{1}{4}\frac{g}{10\pi^{2}}k_{F}^{5}\left|k\right|^{2}}{\frac{1}{4}\frac{g}{6\pi^{2}}k_{F}^{3}}}=\sqrt{\frac{g}{3\pi^{2}}k_{F}^{3}+\frac{12}{5}k_{F}^{2}\left|k\right|^{2}}
\end{equation}
which is the previously mentioned equation \eqref{eq:eps-k-af-thm}.

\subsection*{Further explanation of \eqref{eq:intro-gen-1} for general potentials}

Our analysis can be extended easily to any potential satisfying 
\begin{equation}
\hat V_k \ge 0, \quad \sum_{k\in \mathbb{Z}^{3}_*} \hat V_k^2 <\infty.
\end{equation}
To be precise, let $k\in\mathbb{Z}_{\ast}^{3}$ and $M\in\mathbb{N}$ satisfy $\hat{V}_{k}\gg k_{F}^{-1}$ and $1\le M\ll k_{F}\left|k\right|^{\frac{1}{2}}$, and let $\epsilon_k, \varphi_k$ and $\hat \Psi_M$ be as in Theorem \ref{thm:CoulombPlasmonStates}. The proof of Proposition \ref{prop:SpectralEstimate} remains unchanged and we only need to generalize slightly the one-body estimates in Proposition \ref{prop:ExplicitOneBodyQuantities}. We can use exactly \eqref{eq:varphi-inf-0}, \eqref{eq:need-to-mod-1} and \eqref{eq:need-to-mod-2}, without substituting $\hat V_k= g |k|^{-2}$, to get 
\begin{align} \label{eq:varphi-inf-gen}
\left\Vert \varphi\right\Vert _{\infty} & \leq C\frac{k_{F}^{3}\left|k\right|^{2}\hat{V}_{k}}{k_{F}^{3}\left|k\right|^{2}\hat{V}_{k}-Ck_{F}^{2}\left|k\right|^{2}}\frac{\sqrt{k_{F}\left|k\right|}}{\sqrt{k_{F}^{3}\left|k\right|^{2}\hat{V}_{k}}}\sqrt{\hat{V}_{k}}\nonumber\\
&=C\frac{\hat{V}_{k}}{\hat{V}_{k}-Ck_{F}^{-1}}\frac{1}{\sqrt{k_{F}^{2}\left|k\right|}}  \leq C\frac{1}{\sqrt{k_{F}^{2}\left|k\right|}} 
\end{align}
under the condition that $\hat{V}_{k}\gg k_{F}^{-1}$. Using again 
\eqref{eq:varphi-6-0} and $\sqrt{\left|L_{k}\right|}\sim\sqrt{k_{F}^{2}\left|k\right|}$, we also have
\begin{equation} \label{eq:varphi-6-gen}
\left\Vert \varphi\right\Vert _{6}^{3}=\sqrt{\sum_{p\in L_{k}}\left|\left\langle e_{p},\varphi\right\rangle \right|^{6}}\leq\sqrt{\left|L_{k}\right|}\left\Vert \varphi\right\Vert _{\infty}^{3}  \leq C\frac{1}{k_{F}^{2}\left|k\right|}. 
\end{equation}
Moreover, we can split 
\begin{equation}
\sum_{l\in\mathbb{Z}_{\ast}^{3}}\left\Vert \tilde{E}_{l}-h_{l}\right\Vert _{\mathrm{HS}}^{2}=\sum_{l\in2B_{F}}\left\Vert \tilde{E}_{l}-h_{l}\right\Vert _{\mathrm{HS}}^{2}+\sum_{l\in\mathbb{Z}^{3}\backslash2B_{F}}\left\Vert \tilde{E}_{l}-h_{l}\right\Vert _{\mathrm{HS}}^{2}
\end{equation}
and estimate by Lemma \ref{prop:ExchangeHilbertSchmidtEstimate}
that for $\left|l\right|\leq2k_{F}$
\begin{align}
\left\Vert \tilde{E}_{l}-h_{l}\right\Vert _{\mathrm{HS}}^{2} & \leq\min\left\{ 2\left\langle v_{l},h_{l}v_{l}\right\rangle ,4\left\Vert v_{l}\right\Vert ^{4}\right\} \leq C\min\left\{ k_{F}^{3}\left|l\right|^{2}\hat{V}_{l},k_{F}^{4}\left|l\right|^{2}\hat{V}_{l}^{2}\right\} \\
 & \leq C\min\left\{ 1,k_{F}\hat{V}_{l}\right\} k_{F}^{3}\hat{V}_{l}\left|l\right|^{2}.\nonumber 
\end{align}
For $\left|l\right|>2k_{F}$ we simply estimate
\begin{equation}
\left\Vert \tilde{E}_{l}-h_{l}\right\Vert _{\mathrm{HS}}^{2}\leq4\left\Vert v_{l}\right\Vert ^{4}\leq C\left|L_{l}\right|^{2}\hat{V}_{l}^{2}\leq Ck_{F}^{6}\hat{V}_{l}^{2}.
\end{equation}
Thus
\begin{equation} \label{eq:HS-gen}
\sum_{l\in\mathbb{Z}_{\ast}^{3}}\left\Vert \tilde{E}_{l}-h_{l}\right\Vert _{\mathrm{HS}}^{2}\leq Ck_{F}^{3}\left(\sum_{l\in2B_{F}}\min\left\{ 1,k_{F}\hat{V}_{l}\right\} \hat{V}_{l}\left|l\right|^{2}+Ck_{F}^{3}\sum_{l\in\mathbb{Z}^{3}\backslash2B_{F}}\hat{V}_{l}^{2}\right).
\end{equation}
Inserting \eqref{eq:varphi-inf-gen}, \eqref{eq:varphi-6-gen} and \eqref{eq:HS-gen} in the estimate in Proposition \ref{prop:SpectralEstimate}, we obtain \eqref{eq:intro-gen-1}.

\bigskip

{\bf AUTHOR DECLARATIONS} 

\bigskip

{\bf Conflict of Interest:} The authors have no conflicts to disclose.

\bigskip

{\bf Data Availability:} Data sharing is not applicable to this article as no new data were created or analyzed in this study.

\end{document}